\documentclass[journal]{IEEEtran}
\IEEEoverridecommandlockouts
\usepackage{cite}
\usepackage{amsmath,amssymb,amsfonts}
\usepackage{graphicx}
\usepackage{textcomp}
\usepackage{xcolor}
\usepackage{subfigure}
\usepackage{multirow}
\usepackage{diagbox}
\usepackage{booktabs}
\usepackage{algorithm}
\usepackage{algpseudocode}
\usepackage{amsmath}
\usepackage{threeparttable}
\algrenewcommand\algorithmicindent{0.5em} 

\def\BibTeX{{\rm B\kern-.05em{\sc i\kern-.025em b}\kern-.08em
    T\kern-.1667em\lower.7ex\hbox{E}\kern-.125emX}}
\begin{document}

\title{Is DRL-based MAC Ready for Underwater Acoustic Networks? Exploring Its Practicality in Real Field Experiments}

  \author{Jiani Guo,
      Bingwen Huangfu,
      Shanshan Song*,~\IEEEmembership{Member,~IEEE,}
      Nan Sun,
      Miao Pan,~\IEEEmembership{Senior Member,~IEEE,}
      Guangjie Han~\IEEEmembership{Fellow,~IEEE,}

  \thanks{Shanshan Song is the corresponding author.}
  \thanks{Shanshan Song, Bingwen Huangfu, Jiani Guo, and Nan Sun are with the College of Computer Science and Technology, Jilin University, Changchun 130012, China (e-mail: songss@jlu.edu.cn; hfbw24@mails.jlu.edu.cn; jnguo@jlu.edu.cn; sunnan23@mails.jlu.edu.cn).}%
  \thanks{Miao Pan is with the Department of Electrical and Computer Engineering, University of Houston, Houston, TX 77204 USA (e-mail: mpan2@uh.edu).}
  \thanks{GuangJie Han is with the Department of Internet of Things Engineering, Hohai University, Changzhou 213022, China (e-mail: hanguangjie@gmail.com).}
  }


\maketitle


\maketitle

\begin{abstract}
Medium Access Control (MAC) protocols rely on neighbor and environment information to design collision-free access rules for Underwater Acoustic Networks (UANs). Acquiring this information suffers from high communication overhead due to the unique underwater acoustic channel characteristics, such as long propagation delay, spatiotemporal variations in communication quality, and high attenuation. Deep Reinforcement Learning (DRL) is promising to circumvent the UANs' physical constraints and provide a low-overhead solution for underwater MAC protocols, since it can decide access rules based on real-time observation without extra information exchange. However, the unique underwater acoustic channel characteristics impose significant challenges on observation acquisition, training time, and the balance of multiple reward factors for DRL-based MAC protocols. Most existing methods remain at the theoretical level: (1) they design partial intelligent agents failing to achieve fully autonomous access; (2) they assume unreasonable simulation scenarios, weakening the effects of underwater acoustic channel characteristics on MAC protocols. To enhance the practicality of DRL-based MAC protocols, we first analyze the application challenges of DRL in UANs through real field experiments. Based on the above challenges, we propose a DRL-based MAC protocol that considers observation loss and balances multiple reward factors to achieve efficient Entire Autonomous access in the UAN (EA-MAC). To further explore the feasibility of DRL-based MAC protocols, we implement EA-MAC and other state-of-the-art protocols on underwater acoustic modems and evaluate their performance in real field experiments. Experimental results demonstrate that EA-MAC can adaptively determine the scheduling sequence for each node, enabling high-throughput and fair communication in a straightforward manner for UANs. 
\end{abstract}


\section{Introduction}
Underwater Acoustic Networks (UANs) have attracted significant attention from researchers since they provide essential technical support for various underwater applications, including environmental observation, marine information collection, and resource exploration \cite{r1, r14, r15}. However, UANs suffer from unique underwater acoustic channel characteristics, such as long propagation delay, spatiotemporal varying communication quality, and high attenuation, which make even basic point-to-point communication a significant challenge in UANs. Based on such a condition, Medium Access Control (MAC) protocols become crucial in UANs, as they define point-to-point access rules determining the performance bottleneck.

To avoid complex implementation and packet collisions, traditional underwater MAC protocols generally allocate distinct sending time or frequency to different nodes \cite{r2, r16, r17}. Due to long propagation delay (underwater acoustic velocity is around 1500 m/s) and low bandwidth (ranging from tens of hertz to tens of kilohertz), this division approach further reduces the communication resources of a single node. In addition, underwater nodes are typically equipped with half-duplex acoustic modems, indicating that a node can not send and receive simultaneously. That renders protocols associated with carrier sensing ineffective. Under such conditions, researchers study complicated protocols to improve resource utilization \cite{r18, r19, r20}. These protocols rely on neighbor and environment information to allocate communication resources on demand. However, obtaining this information requires additional information exchanges, which is unaffordable for UANs where communication is unreliable.   

\textit{Can underwater MAC protocols simultaneously adopt a simple implementation scheme and improve resource utilization?} To pursue this objective, researchers have started applying Deep Reinforcement Learning (DRL) to the underwater MAC design. By leveraging the sensing and learning capabilities of DRL, nodes acting as intelligent agents can adaptively determine their sending time while utilizing the full frequency spectrum, achieving collision-free communication without complex information exchanges \cite{r21}. However, most existing DRL-based underwater MAC protocols lack the consideration of real ocean scenarios and remain at the theoretical level, facing the following issues:

\textbf{(1) Partial Intelligent Access.} Most existing protocols design partial nodes as intelligent agents, while other nodes still employ traditional protocols such as random access and Time Division Multiple Access (TDMA) \cite{r5} \cite{r6}. In such a condition, only a few nodes access the UAN autonomously, failing to improve the whole network performance. \textbf{(2) Unrealistic Simulation Conditions.} Some protocols ignore long propagation delay, assuming that rewards can be obtained immediately after an action is executed, which does not correspond to real ocean scenarios. Some protocols consider the long propagation delay of UANs but assume it is an integer multiple of the transmission delay, neglecting the varying propagation delay from different nodes. Moreover, these protocols overlook the trade-off between overall throughput and individual fairness, failing to enhance the UAN performance in real ocean scenarios. 

To fully leverage DRL in designing simple and efficient underwater MAC protocols, we first analyze the effects of underwater acoustic channel characteristics on DRL through real field experiments. Targeting these practical problems, we design an Entire Autonomous MAC protocol (EA-MAC) based on DRL, which considers long propagation delay, high attenuation, and complex factor interactions. In addition, we implement EA-MAC and other state-of-the-art protocols on underwater acoustic modems and evaluate their performance in real field experiments, further exploring whether DRL-based MAC is ready for UANs.

We summarize our contributions as follows:

\begin{itemize}
    \item We organize real field experiments, analyze the effects of underwater acoustic channel characteristics on DRL methods in real ocean environments, and explore the practical issues of DRL methods in underwater applications. These practical problems are summarized as uncertain reward acquisition delay, incomplete observations, and a trade-off between throughput and fairness.     
    \item Based on the above practical problems, we design DRL-based EA-MAC which 1) relaxes the assumption of fixed reward acquisition delay in existing algorithms, aligning more closely with real UANs; 2) employs Bayesian inference to deduce and compensate for missing data transmission observations, enhancing the learning capability of individual nodes; 3) designs a reward function balancing network throughput and transmission fairness among nodes to further improve UAN performance.
    \item Extensive simulation and real field experiments demonstrate that the proposed MAC protocol enhances the adaptability of DRL to UANs, outperforming other DRL-based MAC protocols. Meanwhile, due to the integration of a DRL method, our EA-MAC can achieve collision-free and fair transmission in a simple manner.
\end{itemize}

\section{Related Work}

\subsection{Underwater MAC Protocols without Equipment Constraints} 

Traditional underwater MAC protocols are mostly modified based on fundamental terrestrial wireless network MAC protocols (such as random access-based protocols and TDMA-based protocols). These protocols only rely on the resources of the MAC layer to schedule multiple nodes, achieving network communication. \cite{r7} analyzed the influence of time and space on the underwater Slotted ALOHA MAC protocol and determined the spatiotemporal related interference regions, to improve the successful transmission probability in UANs. In \cite{r8}, the authors designed a TDMA-based protocol, which allocates slots at a packet level to reduce transmission delay. \cite{r9} adopted random time slot scheduling created by seeds to improve the communication efficiency of UANs. After exchanging seeds, nodes obtain the scheduling arrangements of other neighbors and transmit data in the same slot with those that do not cause spatiotemporal conflicts. These protocols are hardware-friendly and easy to implement, making them suitable for deployment across a wide range of underwater acoustic modems. However, random access-based protocols struggle to effectively resolve transmission conflicts, leading to increased retransmission delay, while TDMA-based protocols face low flexibility, resulting in higher queuing delay.

\subsection{Underwater Cross-layer MAC Protocols} 

With the rapid advancement of underwater acoustic communication technology, researchers have begun exploring cross-layer underwater MAC protocols that integrate time resources at the MAC layer with communication resources at the physical layer (such as spectrum, power, and coding) to enable flexible, efficient, and collision-free node access. In \cite{r3}, the authors proposed a Multi-Carrier Code Division Multiple Access (MC-CDMA)-based MAC protocol, which jointly allocates subcarrier-related parameters and transmission power to achieve efficient concurrent communication and optimize throughput, delay, and energy consumption in mobile UANs. \cite{r4} proposed a MAC protocol based on Orthogonal Frequency Division Multiplexing (OFDM) technology, adjusting the number of subcarriers and transmission power by considering dynamic traffic load to improve throughput. In \cite{r13}, the authors proposed a protocol that combines power-domain and code-domain hybrid non-orthogonal multiple access (NOMA) technology to classify senders and allocate communication resources, achieving high-concurrency communication. Although this cross-layer design concept significantly enhances the flexibility of resource allocation and the capability of collision avoidance in UANs, it requires underwater acoustic modems to support corresponding modulation techniques and an extra handshake process to obtain neighbors' information, thereby increasing both implementation and computational complexity.

\subsection{Underwater DRL-based MAC Protocols} 

Based on the above analysis, DRL is promising for UANs to achieve both simple and efficient node access, since it treats the node as a DRL agent, and the agent can learn to find an optimal scheduling strategy without relying on detailed information from additional handshake processes. In this way, DRL-based MAC protocols can rely on a simple slotted random access mechanism to achieve high concurrency performance comparable to that of complex protocols. In \cite{r5}, the authors proposed a DRL-based MAC protocol in which the agent can vary the start time in each transmission time slot to further exploit the spatiotemporal uncertainty of the UANs. \cite{r6} proposed a DON-based MAC protocol that incorporates the long propagation delay into the DRL framework and modifies the DRL algorithm accordingly to find the optimal scheduling strategy for limited intelligent nodes. In \cite{r10}, the authors proposed a deep multi-agent reinforcement learning (MARL) strategy to maximize the number of successful communications while minimizing failures caused by conflicts or interference. Meanwhile, historical observations are leveraged to reduce decision uncertainty by bridging the gap between partial observations and global states.

However, most existing DRL-based MAC protocols remain at the theoretical stage with only partial agents, relying on numerous unrealistic assumptions of the underwater acoustic environment, which limit their practical applicability in real UAN scenarios. To explore the practicality of DRL and enhance UANs' performance, we analyze real-world field experimental results and propose a DRL-based MAC protocol, in which each node acts as an intelligent agent that adaptively determines its own scheduling strategy by considering the unique characteristics of the underwater acoustic channel.  

\begin{figure}[t]
	\centering	
	\subfigure[Test topology\label{1a}]{\includegraphics[width=4cm,height=5cm]{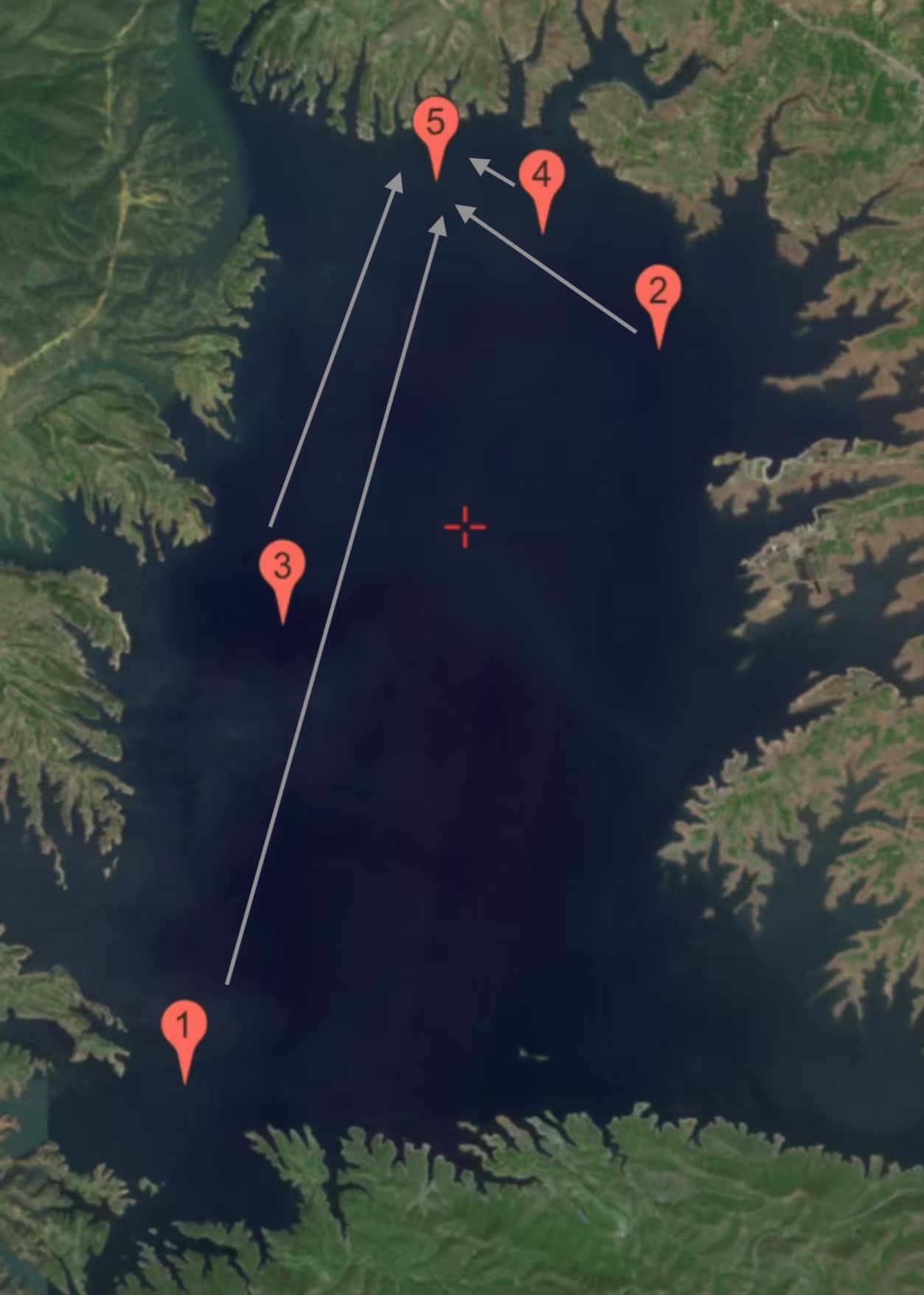}}\hspace{0.1em}  
	\subfigure[Test equipments
	\label{1b}]{\includegraphics[width=4cm,height=5cm]{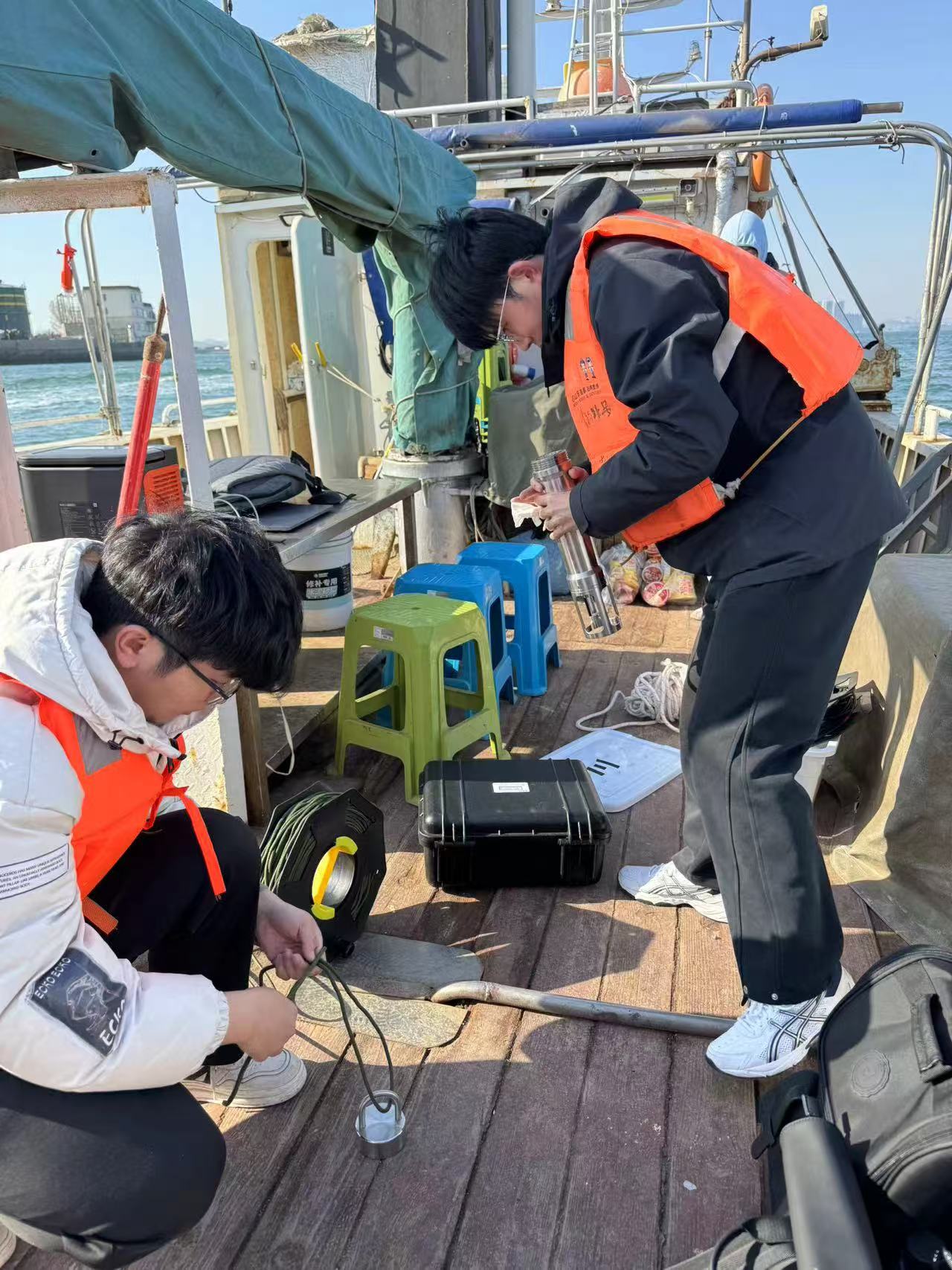}}	
    
	\subfigure[Test ships
	\label{1c}]{\includegraphics[width=8cm]{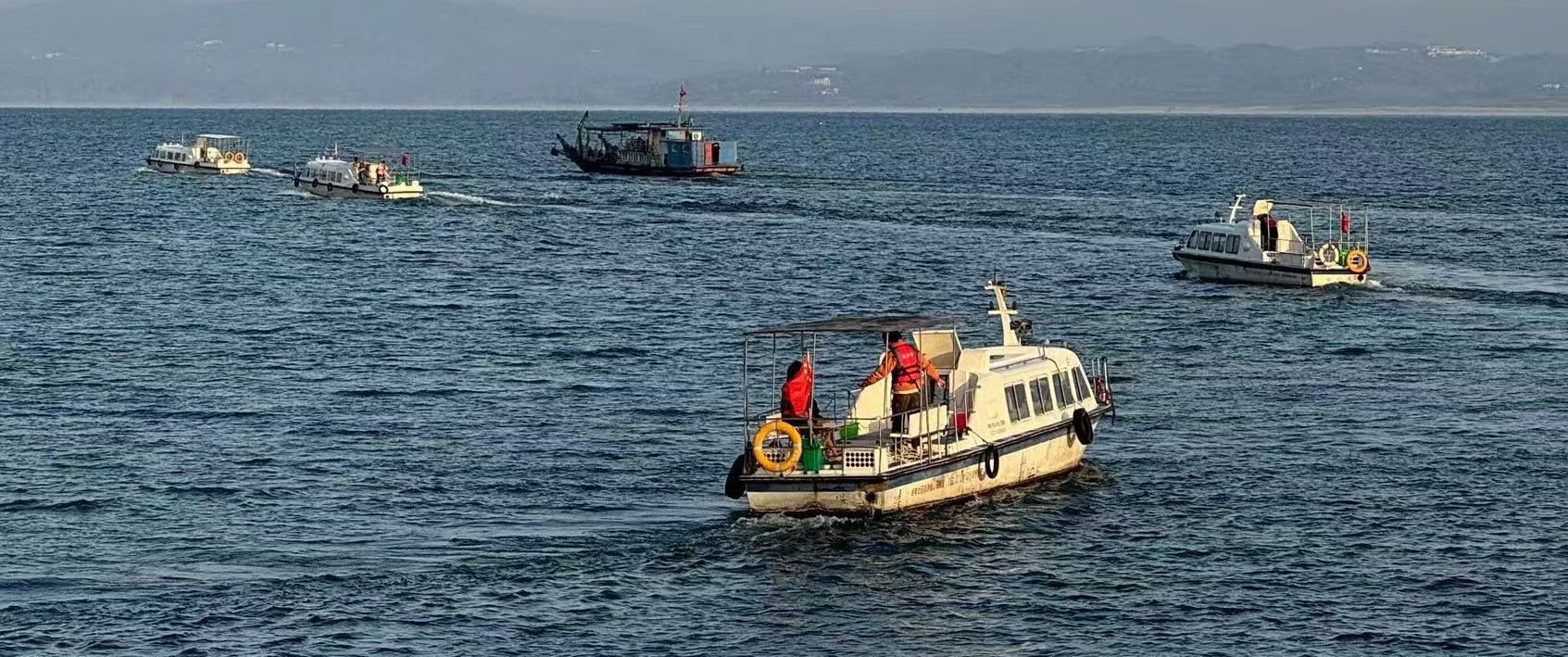}}	
	\caption{A 5-node UAN deployed in
Danjiangkou Reservoir, Henan, China, 2024. Node~1 to node~4 are senders, and node~5 is the sink node.} 
	\label{1}
\end{figure} 

\section{Field Observations \label{fo}}

To explore the effects of underwater acoustic channel characteristics on DRL-based MAC protocols, we deployed a 5-node UAN in the Danjiangkou Reservoir in 2024, as represented in Fig.~\ref{1}. Nodes 1 to 4 adopt the TDMA protocol and send packets to node 5 in their assigned slots. After receiving a data packet, node 5 replies with an acknowledgment (ACK) packet in the current slot to the sender. In such a scenario, we observe UAN communication results from different perspectives and summarize the following problems that affect the application of DRL in underwater MAC protocols.

\begin{figure}[t]
  \centering
  \includegraphics[width=8cm]{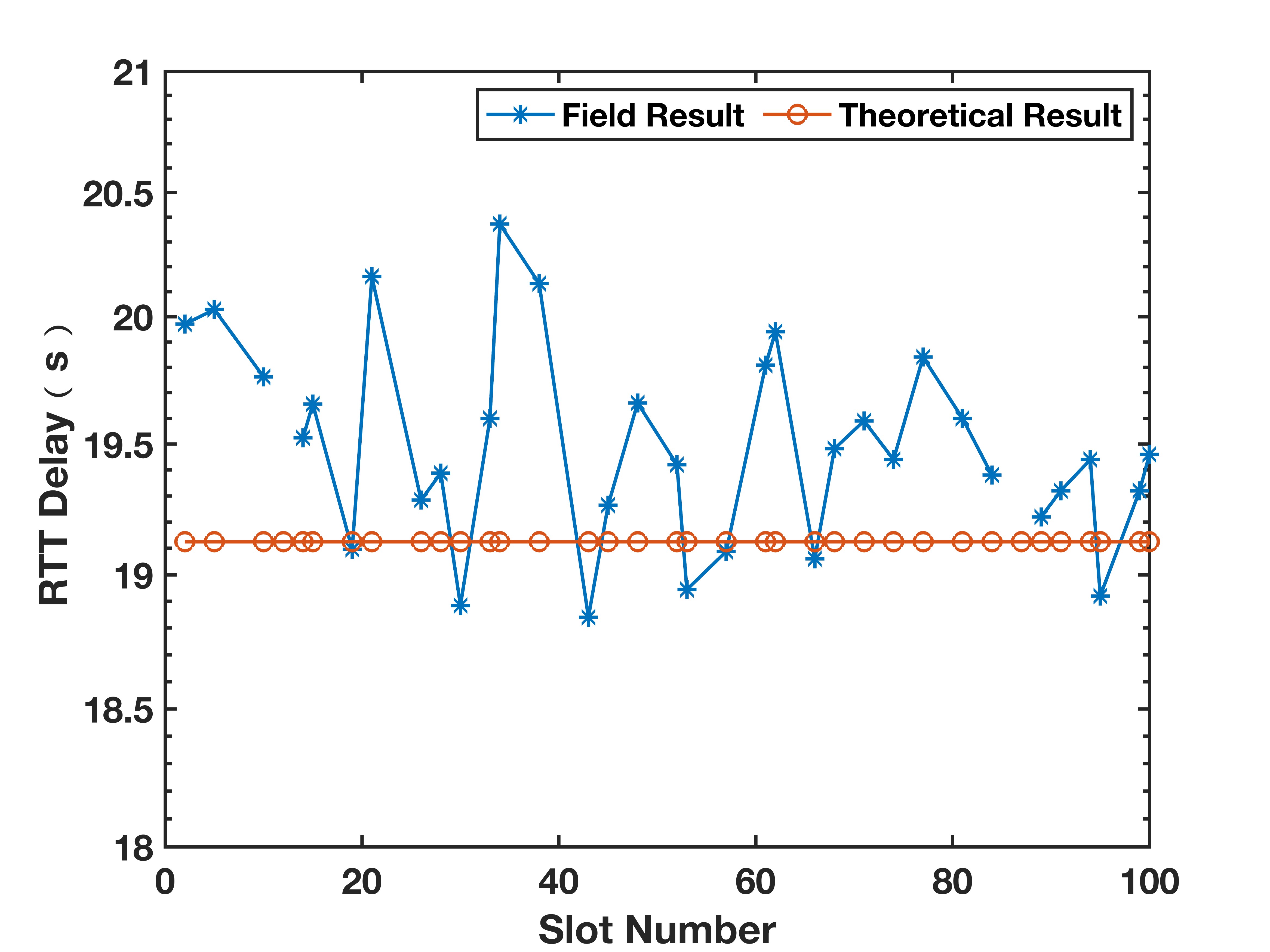}
  \caption{Comparison of Round-Trip Time (RTT) delay between field experiments and theoretical value. RTT delay means the time from sending a data packet to receive the ACK packet. The absence of markers indicates packet loss. Theoretical Result is the twice propagation delay.}
  \label{2}
\end{figure}

\textbf{Uncertain reward acquisition delay.} DRL-based MAC protocols for UANs typically use successful transmission counts as the primary basis for reward to enhance network performance.  Therefore, senders as agents rely on ACK packets to obtain reward information since receiving a correct ACK packet indicates that the transmission is successful. However, existing protocols ignore long propagation delay or are limited by the simulation capabilities, assuming 1) senders can receive an ACK packet immediately after sending a data packet; 2) the propagation delay is an integer multiple of the transmission delay; or 3) the ACK acquisition delay is twice the propagation delay. These assumptions do not align with our field results. As represented in Fig.~\ref{2}, we take node~1 as an example, recording its delay from sending a data packet to receiving the ACK packet over a period of time. We refer to the delay as Round-Trip Time (RTT), which represents the reward acquisition delay. Based on the recording results, we can observe that RTT is variable, indicating that it does not exhibit an integer multiple relationship with either propagation delay or transmission delay. To enhance the practicality of underwater DRL-based MAC protocols, such uncertain reward acquisition delay should be considered in the training process.

\begin{table}[t]
\caption{Successful reception ratio among different nodes}
\begin{tabular}{llllll}
\toprule
\diagbox[width=6.2em]{Receiver}{Sender}   & Node 1 & Node 2 & Node 3 & Node 4 & Node 5 \\ \midrule
Node 1 & -      & 95\%   & 69\%   & 24\%   & 100\%  \\
Node 2 & 94\%   & -      & 81\%   & 29\%   & 96\%   \\
Node 3 & 47\%   & 93\%   & -      & 59\%   & 35\%   \\
Node 4 & 89\%   & 30\%   & 42\%   & -      & 100\%  \\
Node 5 & 94\%   & 64\%   & 56\%   & 39\%   & -      \\ \bottomrule
\end{tabular}
\label{t1}
\end{table}

\textbf{Incomplete observations.} In DRL-based MAC protocols, multiple senders, acting as collaborators, should observe each other's transmission conditions to avoid collisions and enhance throughput performance. Therefore, for a given sender, observing the data transmission status of other senders is crucial for effectively training the scheduling strategy. To evaluate the quality of observations, the UAN is designed as a full-connection network (every two nodes are within each other's communication range) in our real field experiment. We record the successful reception ratio among different nodes over a period of time, and the results are presented in Table~\ref {t1}. Due to high attenuation and spatiotemporal variability of underwater acoustic channels, data loss occurs across most nodes and exhibits directional asymmetry on the same communication link. In such a scenario, senders as agents fail to obtain complete observation information. Existing DRL-based MAC protocols, which ignore severe data loss, cannot be applied to the actual underwater environment.

\begin{figure}[t]
	\centering	
	\subfigure[Data transmission situation of each node under a throughput-only reward model\label{3a}]{\includegraphics[width=4.2cm]{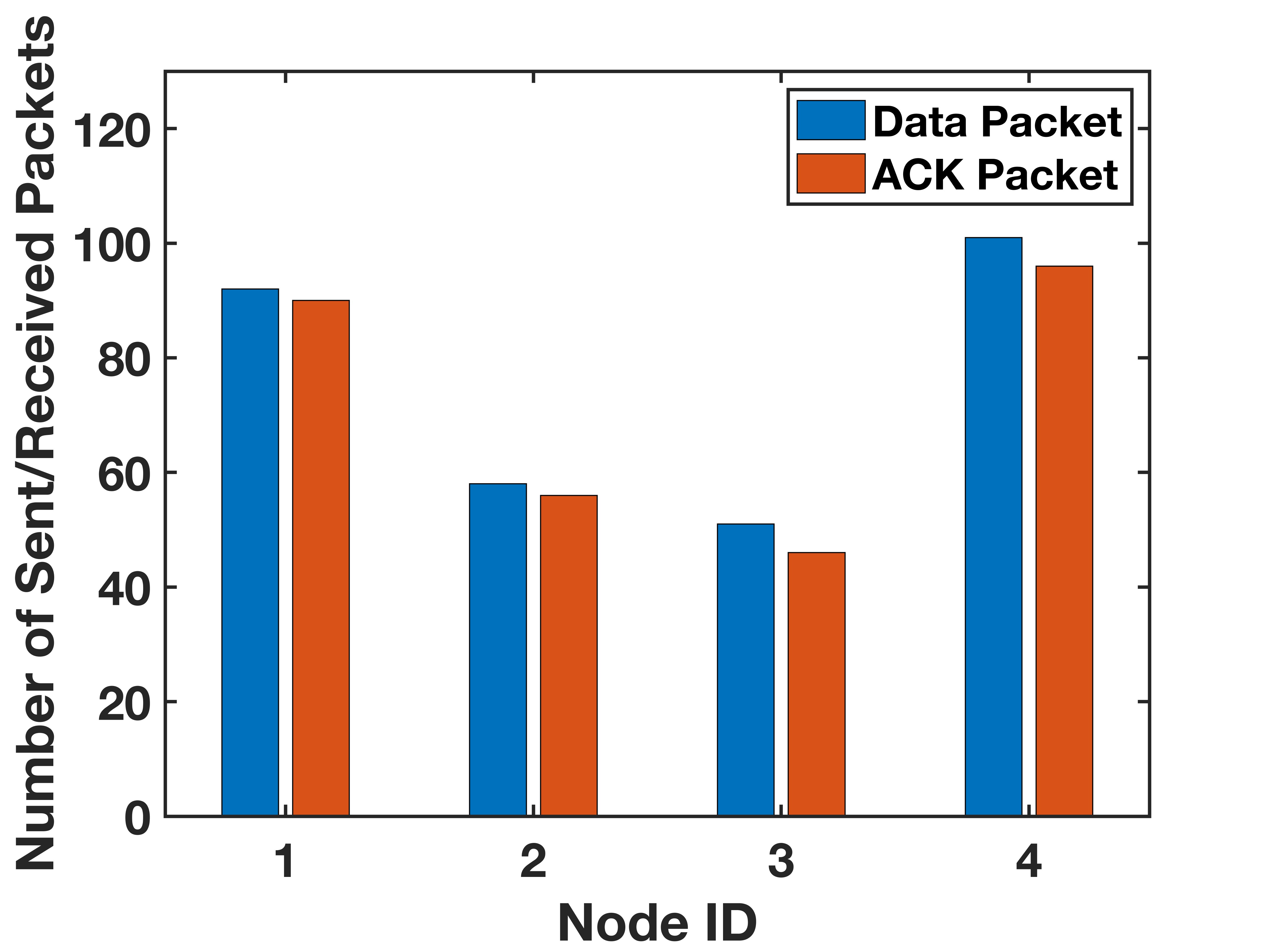}}    
	\subfigure[Data transmission situation of each node under a punishment-only fairness model
	\label{3b}]{\includegraphics[width=4.2cm]{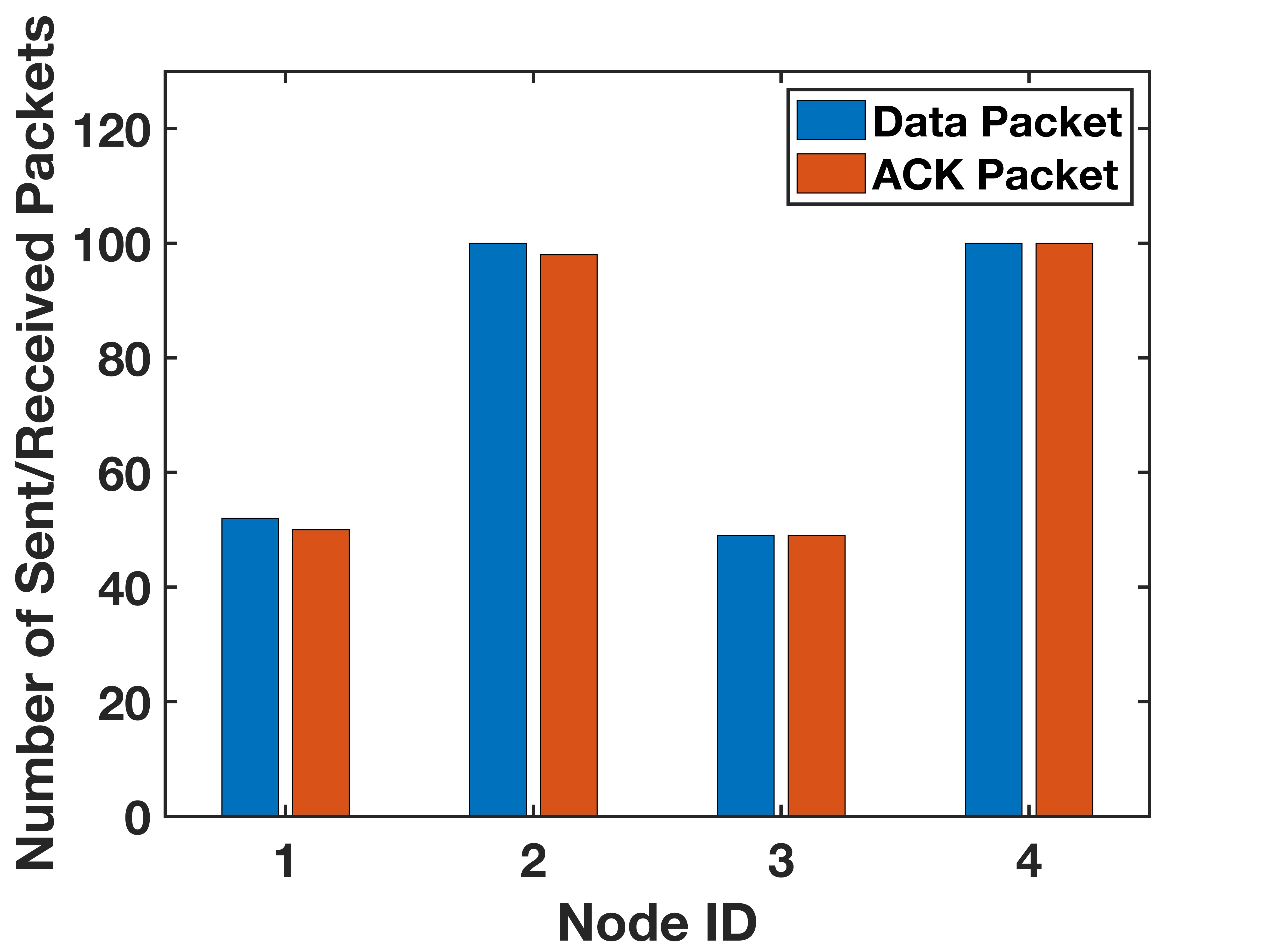}}	
    
	\caption{A real-world problem of balancing throughput and transmission fairness.} 
	\label{3}
\end{figure}

\textbf{Trade-off between throughput and fairness.} In this set of experiments, instead of using the TDMA protocol, we employ a simple Deep Q-Network (DQN) that allows each sender to autonomously decide its own transmission slot. We perform offline training of the DQN based on the actual UAN deployment scenario. After that, we deploy the trained models on different nodes and test them in field experiments. As represented in Fig.~\ref{3a}, the first experiment's reward is designed solely based on the senders' ACK reception count, aiming to maximize throughput. Most existing protocols adopt a similar reward design. However, under such a condition, we can observe significant disparities in the number of packets sent by each node, indicating an unfair distribution of transmission opportunities among nodes. Based on the design of experiment 1, we introduce a suppression penalty for nodes with excessive transmission attempts, as shown in Fig.~\ref{3b}. Field results indicate that such a monotonic penalty fails to effectively address the issue of transmission fairness among nodes. Existing DRL-based MAC protocols typically overlook the trade-off between throughput and fairness, failing to achieve high-performance node scheduling in real-world UANs.

\section{Design of the EA-MAC}

We design EA-MAC based on a mechanism similar to slotted ALOHA. In each slot, nodes utilize a DQN-based algorithm to adaptively decide their action. Due to simple and sparse UAN structures, a distributed DQN framework significantly reduces algorithmic complexity compared to a Multi-Agent Reinforcement Learning (MARL) framework, as each node independently learns its transmission policy without requiring coordination or policy sharing among agents. Meanwhile, leveraging the broadcast nature of the channel, nodes can acquire information from others by overhearing, enabling a certain degree of implicit collaboration and information sharing. In addition, using a distributed DQN framework allows for simple and rapid retraining when network dynamics occur. The above advantages make it well-suited for deployment in resource-constrained UANs, meeting the robustness and ease-of-maintenance demands of real-world systems.

Specifically, as shown in Fig.~\ref{5}, we consider a cluster UAN, which is a classical topology widely applied in various underwater scenarios. By leveraging local information and overheard data from neighboring nodes and the sink node, each sender considers unique underwater acoustic characteristics, deduces and completes the possibly lost observation information, and balances the trade-off between throughput and fairness to determine whether to send or not in the current slot. By exploiting the long propagation delay, EA-MAC allows multiple collision-free nodes to transmit data in the same slot without collisions. Therefore, we design an aggregated ACK mechanism, which combines ACK information into a bit sequence to confirm all transmissions in the current slot with low channel overhead. Based on the above descriptions, our DQN-based algorithm are designed as follows:

\begin{figure*}[t]
  \centering
  \includegraphics[width=16cm]{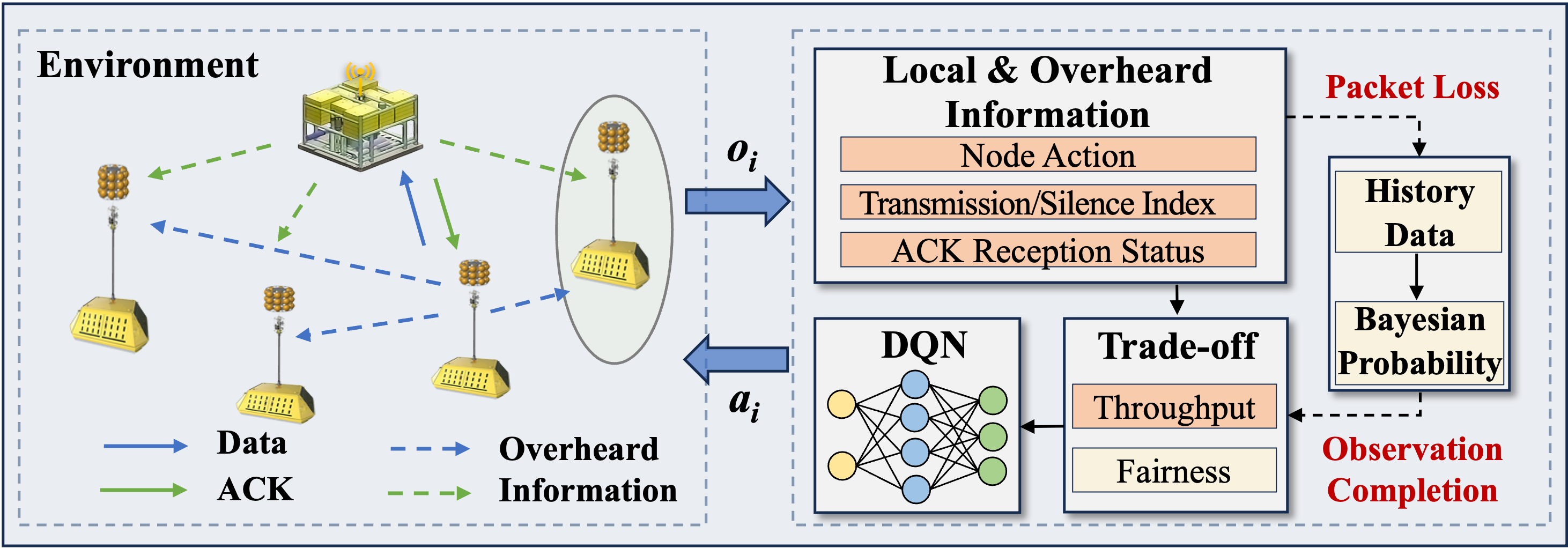}
  \caption{The overview of EA-MAC. }
  \label{5}
\end{figure*}

\begin{figure}[t]
  \centering
  \includegraphics[width=8cm]{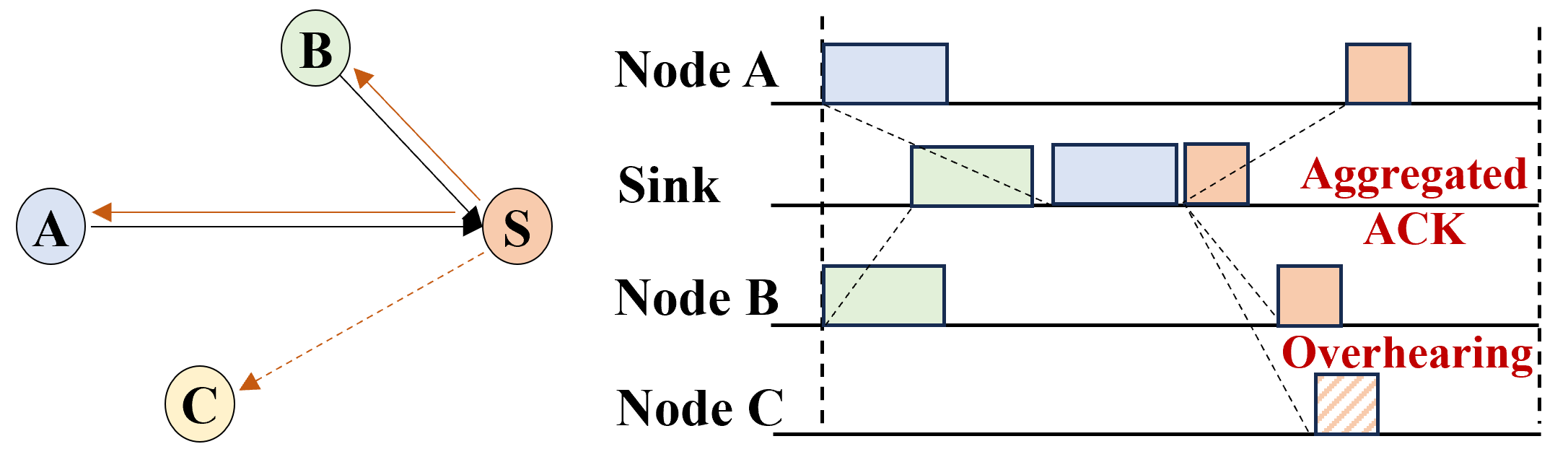}
  \caption{Nodes in UANs utilize long propagation delay, achieving concurrent transmission in the same slot. Sink node acknowledges multiple data using an aggregated ACK.}
  \label{6}
\end{figure}

\textbf{Action and Observation Space}
In each time slot $t$, the EA-MAC node determines whether to access the channel or not. 
We define the action space of the EA-MAC node as:
\begin{equation}
    a^i_t = \{1, 0\},
\end{equation}
where $a^i_t = 1$ means the node $i$ chooses to transmit data in time slot $t$, and $a^i_t = 0$ means the node remains idle.
After executing action at in time slot $t$, the EA-MAC node will obtain the observation $o^(t+1)$ at the end of the slot, which is demonstrated as:
\begin{equation}
    o^i_{t+1} = \{o^i_{action}, o^i_{data}, o^i_{ack}\},
\end{equation}
where $o^i_{action}$ denotes the node's action, $o^i_{data}$ represents the overheard data packets from neighboring nodes, and $o^i_{ack}$ refers to the aggregated ACK information received from the sink node.
However, due to the uncertainty and partial observability in underwater environments, observation within a single time slot may be insufficient to capture the temporal dynamics of nodes' behavior.
To release the delay constraints on observation acquisition and enhance the robustness of decision making, we construct observations over a sliding window of $M$ time slots.
However, an excessively large $M$ can expand the observation space, increasing the complexity of training and execution.

\textbf{Adaptive Suppression-Promotion Fairness.} Based on the analysis in Section \ref{fo}, transmission suppression alone is insufficient to achieve fairness among nodes. To address this problem, the proposed fairness index is designed to simultaneously motivate under-active nodes to transmit and suppress over-active nodes to reduce their transmission frequency. 

We define $N^i_{t}$ as the number of senders confirmed by the broadcasting aggregated ACK received by node $i$ in the current slot. 
The larger $N^i_{t}$ means more concurrent senders. 
Although under-active nodes need to improve sending frequency, it is essential to reduce their interference with the existing high throughput. 
Meanwhile, we must suppress the sending frequency of over-active nodes to provide more opportunities for under-active nodes.
Therefore, we define the fairness penalty as follows.
First, we define the fairness penalty of node $i$ in a single step as:
\begin{equation}
    f_i^t = \begin{cases}
        p / N^i_{t}, & \text{if } a^i_t = 0;\\
        p, & \text{if } a^i_t = 1,
    \end{cases}
    \label{sfairness}
\end{equation}
where $p$ is the base penalty constant.

We define $h^i_t$ as the number of consecutive times that node $i$ selects the same action in the slot $t$, as:
\begin{equation}
    h^i_t = \begin{cases}
        h^i_{t-1} + 1, & \text{if } a^i_t = a^i_{t-1};\\
        0, & \text{otherwise}.
    \end{cases}
\end{equation}
The fairness penalty of node $i$ in slot $t$ is the cumulative fairness penalty over the past $h^i_t$ slots, demonstrated as:
\begin{equation}
    F_i(t) = \sum_{j=0}^{h^i_t} f_i(t-j).
    \label{cfairness}
\end{equation}
The fairness penalty $F_i$ gradually increases with consecutive actions of the node, thereby suppressing over-active nodes and promoting under-active nodes.

\begin{figure}[t]
  \centering
  \includegraphics[width=8cm]{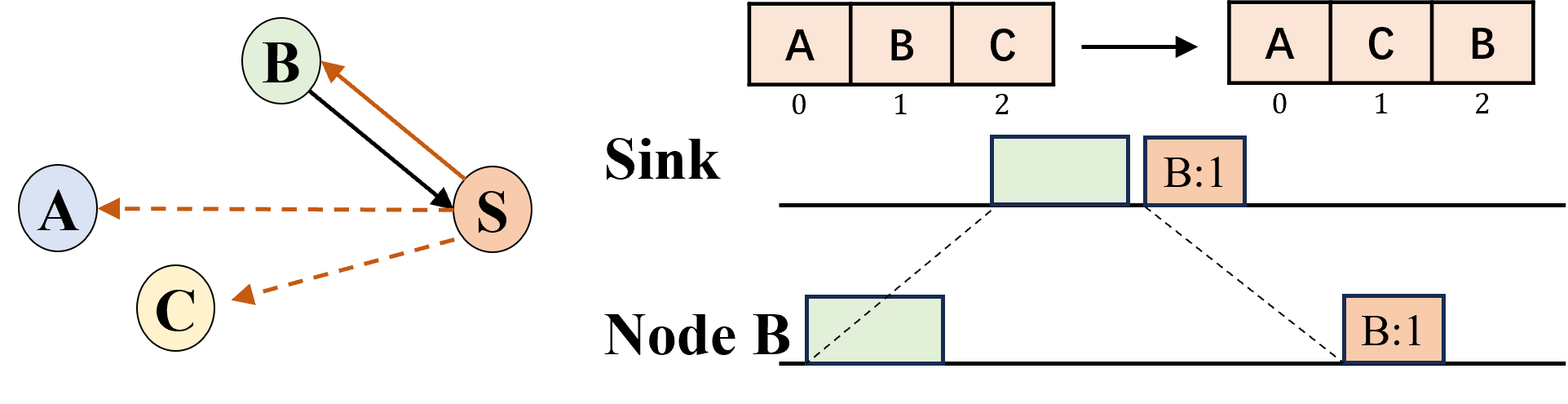}
  \caption{Sink node maintains a transmission node IDs queue. The ACK generated by the sink carries the position of the sender’s ID.}
  \label{order}
\end{figure}

However, despite the ability to overhear data and aggregated ACKs to infer the transmission status of neighboring nodes, 
EA-MAC also faces challenges in coordinating channel competition among nodes due to the decentralized nature of the DQN framework.
We further enhance the aggregated ACK broadcast mechanism to assist EA-MAC in implicit channel access coordination.
Specifically, the sink node maintains a queue $q_{tx}$, initialized with all transmission node IDs in order, as shown in Fig.~\ref{order}.
Upon receiving a data packet, the sink node identifies the position of the sender ID $i$ in $q_{tx}$, demonstrated as $\phi_i$, and moves it to the end of the queue.
Then the sink node encodes the position information $\phi_i$ into the ACK packet and broadcasts it.
We define the transmission order correction index $C_i(t)$ as:
\begin{equation}
    C_i(t) = \begin{cases}
        1- w \cdot \phi_i, & \text{if } a^i_t = 1 \text{ and ACK received};\\
        1, & \text{otherwise},
    \end{cases}
\end{equation}
where $w$ is a weight factor to adjust the impact of the transmission order correction index, which is typically negatively correlated with the number of nodes.
The transmission order correction index $C_i(t)$ regulates the reward based on $\phi_i$, providing an implicit scheduling order for channel access coordination among nodes, thereby reducing collision probability.

\textbf{Reward Function.} For a sender $i$, if $a^i_t=1$ and it successfully receives the corresponding ACK, its reward is defined as $R(t)= C_i(t)\cdot(R_s-F_i(t))$. If $a^i_t=1$ and it fails to receive the corresponding ACK, its reward is defined as $R(t)=R_p-F_i(t)$. $R_s$ denotes a positive reward, while $R_p$ corresponds to a negative penalty. If $a^i_t=0$, the reward only considers the fairness index, denoted as $R(t)=-F_i(t)$. 
The whole training process of EA-MAC is shown in Algorithm.~\ref{eamac}

\begin{algorithm}[t]
\caption{EA-MAC Training Process}
\label{eamac}
\begin{algorithmic}[1]
\State Initialize Q-networks $Q_i(o, a|\theta_i)$, target networks $Q'_i$, and replay buffer $\mathcal{D}_i$ for each node $i$;
\State Initialize sink's transmission queue $q_{tx}$.

\For{each episode}
  \For{each time slot $t$}
    \ForAll{nodes $i = 1$ to $N$ \textbf{in parallel}}
      \State Observe $o^i_t$ and select $a^i_t \in \{0,1\}$ via $\epsilon$-greedy from $Q_i$, then take $a^i_t$ to interact with the environment.
    \EndFor

    \State Sink node receives data, updates $q_{tx}$ and broadcasts aggregated ACK with $\phi_i$.

    \ForAll{nodes $i = 1$ to $N$ \textbf{in parallel}}
      \State Update consecutive action count $h^i_t$;
      \State Compute cumulative fairness penalty $F_i(t)$ using (\ref{sfairness}) and (\ref{cfairness});
      \State Parse $\phi_i$ from ACK (if any);
      \State Compute transmission order correction index $C_i(t)$;
      \State Compute reward $r^i_t$  according to $a^i_t,~F_i(t),~C_i(t)$;
      \State Generate next observation $o^i_{t+1}$;
      \State Sample minibatch from $\mathcal{D}_i$ and update Q-network;
      \State Periodically update target network: $\theta_i^- \leftarrow \theta_i$.
    \EndFor
  \EndFor
\EndFor
\end{algorithmic}
\end{algorithm}

\textbf{Observation Completion.} During the actual testing phase, we consider data loss caused by both collisions and high packet errors in real UANs, completing incomplete observation data. There are several data loss conditions. 

\begin{itemize}
    \item \textbf{Data loss but ACK received:} this condition means a node fails to overhear a neighbor's information. Still, it can deduce the neighbor's action based on the ACK packet to complete its observation.
    \item \textbf{Data loss and ACK loss:} the current node has $N$ neighbors, but it only received $n$ neighbors' information. In such a scenario, we can not determine whether other $N-n$ nodes did not send data or if the data was lost. Therefore, we utilize historical data during the convergence phase of the training process to deduce the action probability of other $N-n$ nodes as follows:  

\begin{align}
&P(a_{i_{n+1}}, \dots, a_{i_N} \mid a_{i_1}, \dots, a_{i_n})= \notag\\ &\frac{P(a_{i_1}, \dots, a_{i_n} \mid a_{i_{n+1}}, \dots, a_{i_N})  P(a_{i_{n+1}}, \dots, a_{i_N})}{P(a_{i_1}, \dots, a_{i_n})},
\label{eq1}
\end{align}
where $a_{i_n}$ determines the $n$-th node's action. It is worth noting that there is no data loss during the training process, and during the convergence phase, the neural network has already learned the conflict relationships between nodes. Therefore, we choose the data for this stage to deduce the unknown information. Finally, we select the action combination with the highest probability to supplement the unknown actions and ACK information.
\item   \textbf{ ACK loss:} a node sends a data packet but fails to receive an ACK packet. In such a scenario, we can not determine whether the data conflicts with other data or the ACK is lost.  The derivation method is the same as that of condition~2. 
\item \textbf{ Continuous ACK loss:} If a node fails to receive ACK for a continuous period of time, we determine that the link between this node and the sink node is disconnected. The node enters a silence state until it re-overhears the sink node's information. 
\end{itemize}

\section{Performance Evaluation}

\subsection{Experiment Settings}

Due to the limitations in the number of experimental modems and constraints of the marine environment, it is not feasible to evaluate the proposed protocol entirely through field experiments. Therefore, this paper adopts a combination of simulation and field experiments to comprehensively evaluate the performance of EA-MAC.

Our underwater acoustic modems operate at 21k~Hz-31k~Hz, with a transmission rate of 100~bps, a transmission power of 30~W, and a state transition time between two data transmissions of around 5~s. The simulation parameters are configured faithfully based on the aforementioned real-world modems. Meanwhile, we employ the fourth-generation ns3-based simulator for UWSNs, called Aqua-Sim~Fg, to conduct simulation experiments. Aqua-Sim~Fg is compatible with various programming languages, including MATLAB, C++, and Python, which provides a general environment to simulate communication technologies, network protocols, and AI models simultaneously. Moreover, Aqua-Sim~Fg can integrate real marine information to simulate underwater signal propagation, packet errors, and packet collisions based on the event sequence. \cite{r11} has proved that Aqua-Sim~Fg is close to real sea trials. In addition, other EA-MAC parameters are summarized in Table.~\ref{t2}. 

From the perspective of MAC protocols, UANs have two types: \textbf{1) non-equidistant UAN} - in such networks, some senders have different propagation distances and can exploit the long propagation delay to transmit packets within the same time slot without collisions, as illustrated in Fig.~\ref{6}; \textbf{2)~equidistant UANs} – in such networks, senders have similar propagation distances, and selecting the same time slot can result in data collisions. The following experiments will consider both types of UANs to evaluate EA-MAC's performance.

\begin{table}[t]
\center
\caption{Successful reception ratio among different nodes}
\begin{tabular}{ll}
\toprule
Parameters                          & Values                  \\ \midrule
Packet size                        & 10 B                   \\
Slot length                        & 9 s                    \\
Number of hidden layers            & 3                      \\
Neuron number of each hidden layer & (128, 256, 128)        \\
Learning rate                      & $10^{-4}$ \\
Batch size                         & 256                    \\
Episodes length                    & 1000$\sim$1800         \\
Maximal steps per episode          & 150                    \\ \bottomrule
\end{tabular}
\label{t2}
\end{table}

\subsection{Simulation Experiments}

In this set of experiments, we vary the number of UANs' nodes from 3 to 6. Real UANs are generally of limited scale, with even their largest practical deployments typically involving only around ten to several tens of nodes. From the perspective of MAC protocols, the receiver is associated with a limited number of senders. Therefore, the setting is reasonable for sparse UANs. For non-equidistant UANs, senders are randomly deployed around the sink at a distance of 1~km~$\sim$~5~km. For equidistant UANs, all senders are positioned uniformly at a fixed distance of 5~km from the sink.   
\begin{figure}[t]
	\centering	
	\subfigure[Non-equidistant UANs' reward results\label{7a}]{\includegraphics[width=4.2cm]{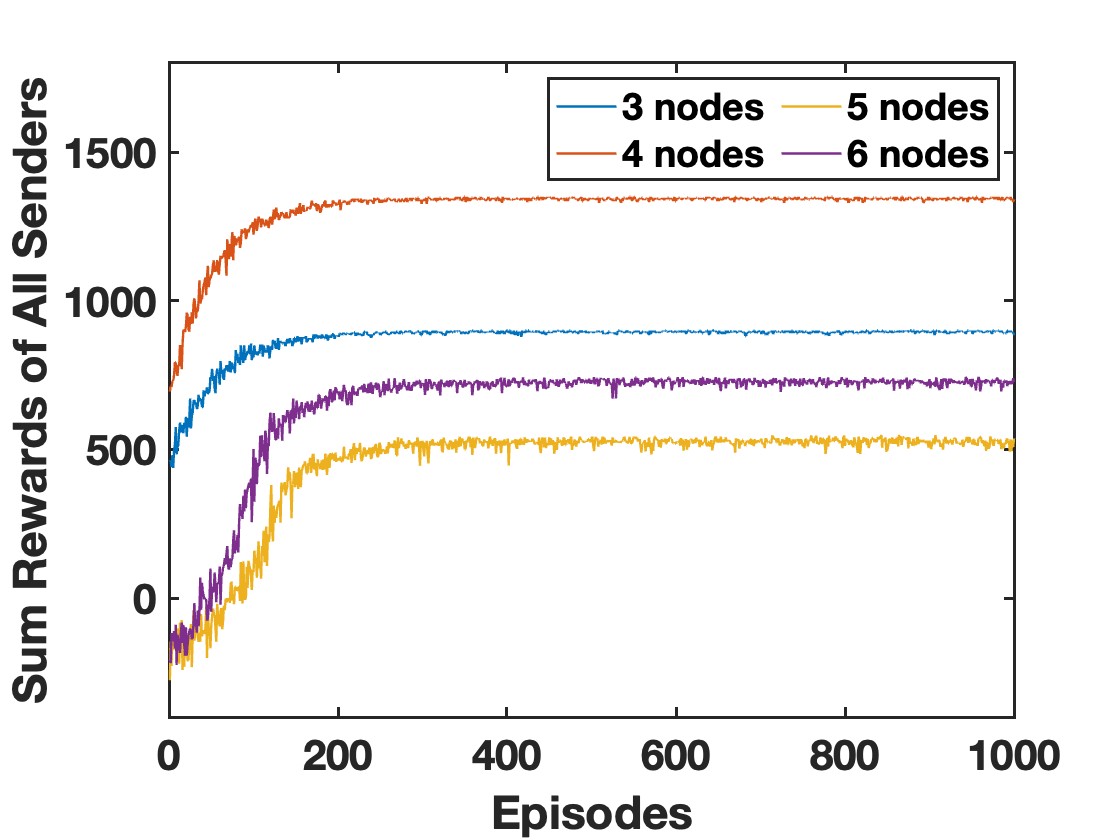}}    
	\subfigure[Equidistant UANs' reward results
	\label{7b}]{\includegraphics[width=4.2cm]{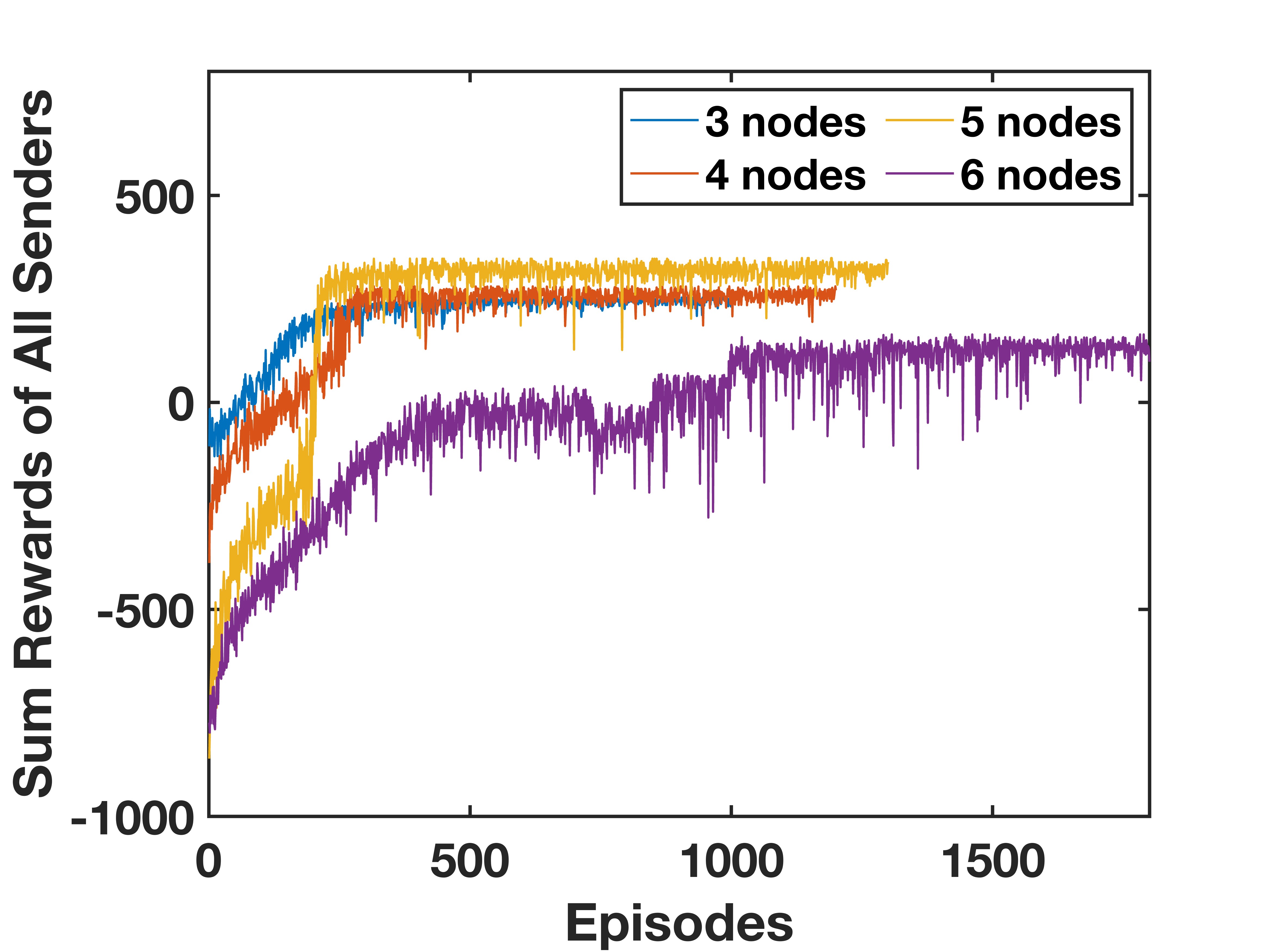}}	    
	\caption{The convergence of EA-MAC during training.} 
	\label{7}
\end{figure} 

\begin{figure}[t]
	\centering	
	\subfigure[Non-equidistant UANs \label{8a}]{\includegraphics[width=4.2cm]{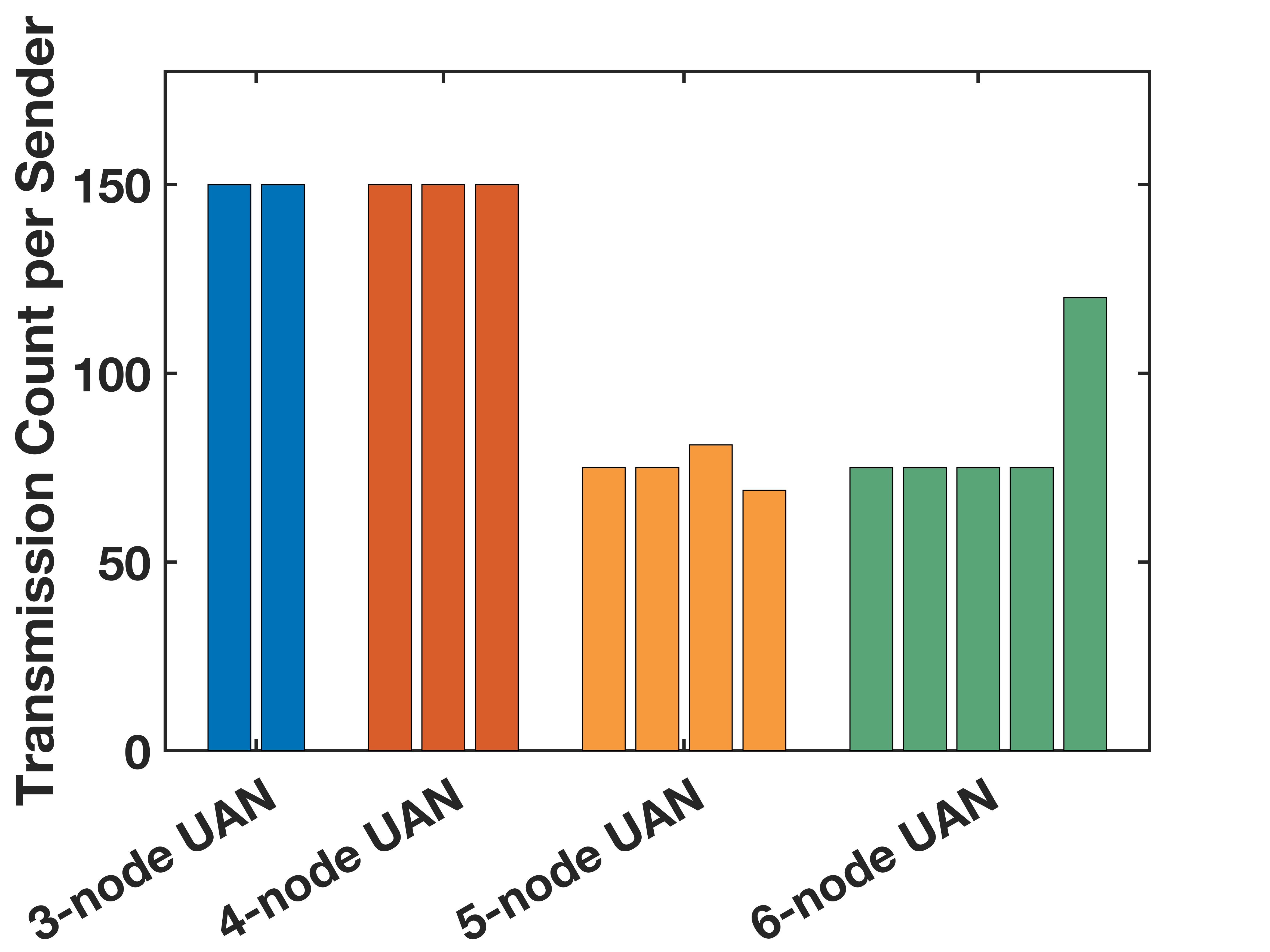}}    
	\subfigure[Equidistant UANs
	\label{8b}]{\includegraphics[width=4.2cm]{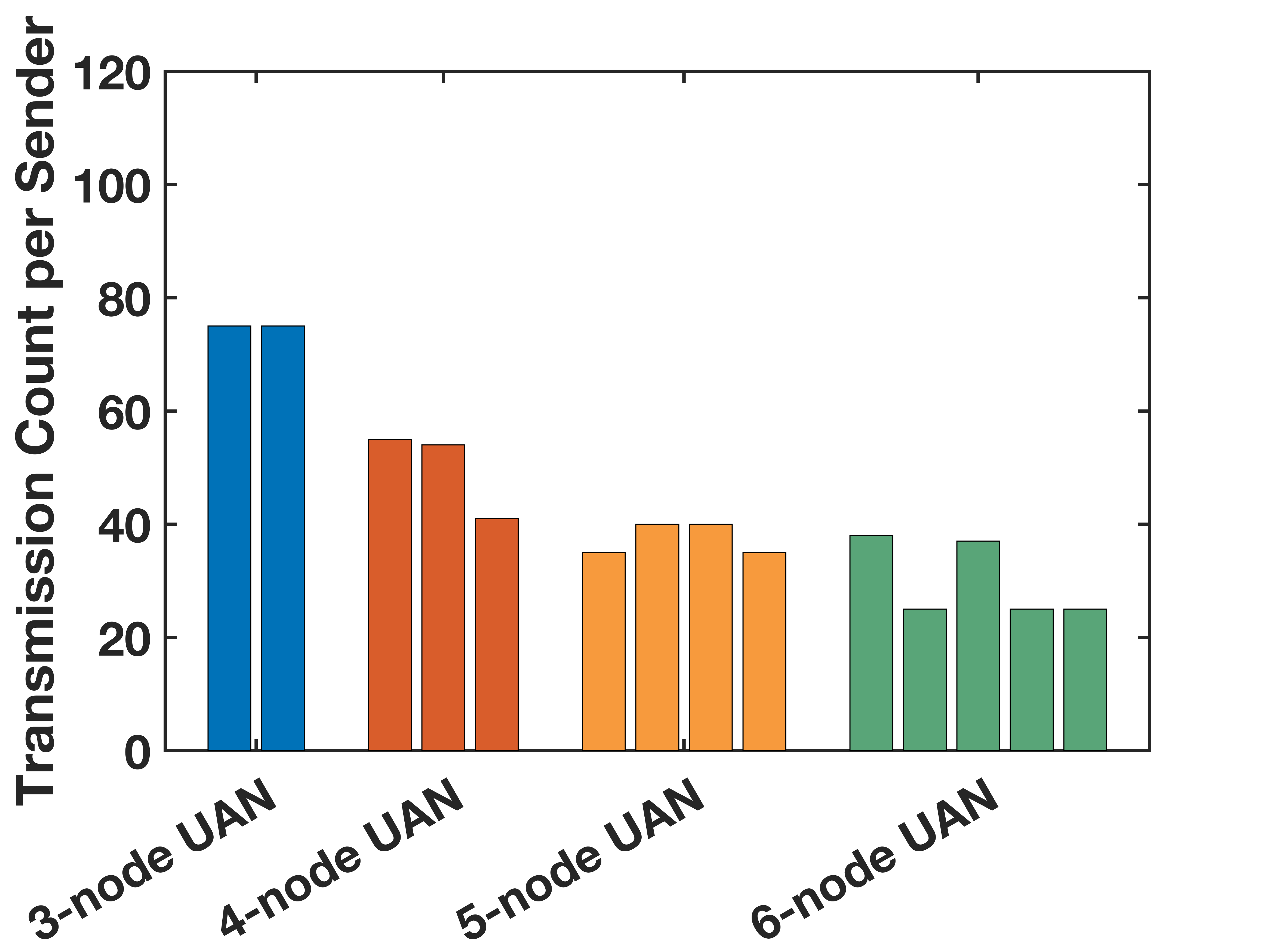}}	    
	\caption{The number of data packets sent by each sender in different UANs.} 
	\label{8}
\end{figure} 

\subsubsection{Convergence} We simulate the convergence of EA-MAC as shown in Fig.~\ref{7}. Simulation results show that the algorithm converges more easily and achieves a higher sum reward in non-equidistant UANs compared to equidistant UANs. This is because collision conditions are more complex in equidistant UANs, making it more challenging to balance the trade-off between throughput and fairness. Such differences in convergence also indicate that EA-MAC can adaptively adjust node access strategy to accommodate varying scenarios, enabling efficient and fair collision-free access.
\begin{table}[t]
\center
\caption{ Trade-off of throughput and fairness in different transmission conditions }
\begin{tabular}{lllllllll}
\toprule
& \multicolumn{4}{l}{Transmission Count} & \multicolumn{4}{l}{Received ACK Count} \\ \cmidrule(lr){2-5} \cmidrule(lr){6-9} 
Node Index                                                         & 1        & 2       & 3       & 4       & 1       & 2       & 3       & 4       \\ \midrule
No Packet loss                                                    & 35       & 40      & 40      & 35      & 35      & 40      & 40      & 35      \\
Packet loss                                                       & 46       & 45      & 50      & 41      & 18      & 16      & 11      & 18      \\
\begin{tabular}[c]{@{}l@{}}Observation \\ completion\end{tabular} & 35       & 40      & 40      & 35      & 29      & 31      & 29      & 23      \\ \bottomrule
\end{tabular}
\label{t3}
\end{table}

\begin{figure}[t]
	\centering	
	\subfigure[UANs without packet loss \label{9a}]{\includegraphics[width=9cm]{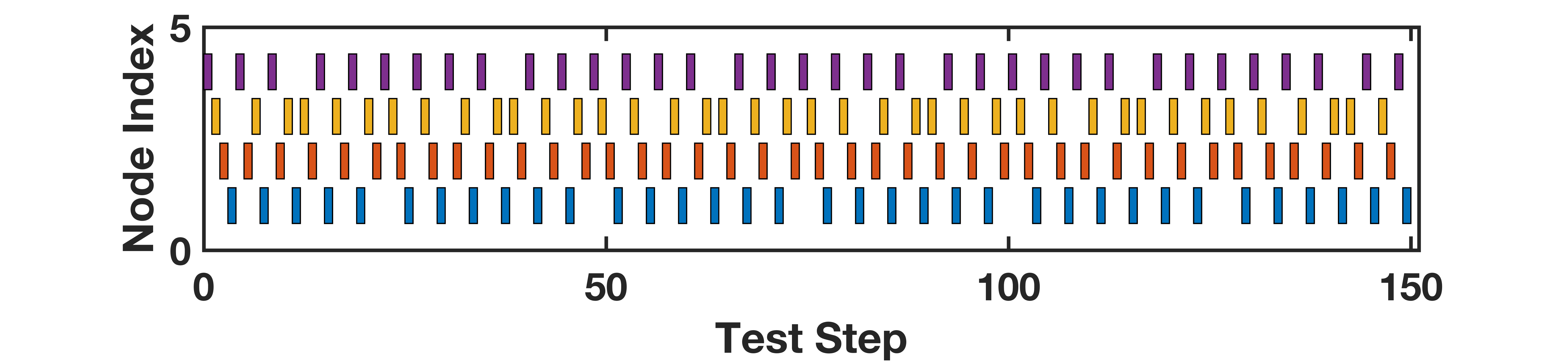}}  
    \vspace{-0.3cm}
	\subfigure[UANs loss 20\% data packets and 30\% ACK packets
	\label{9b}]{\includegraphics[width=9cm]{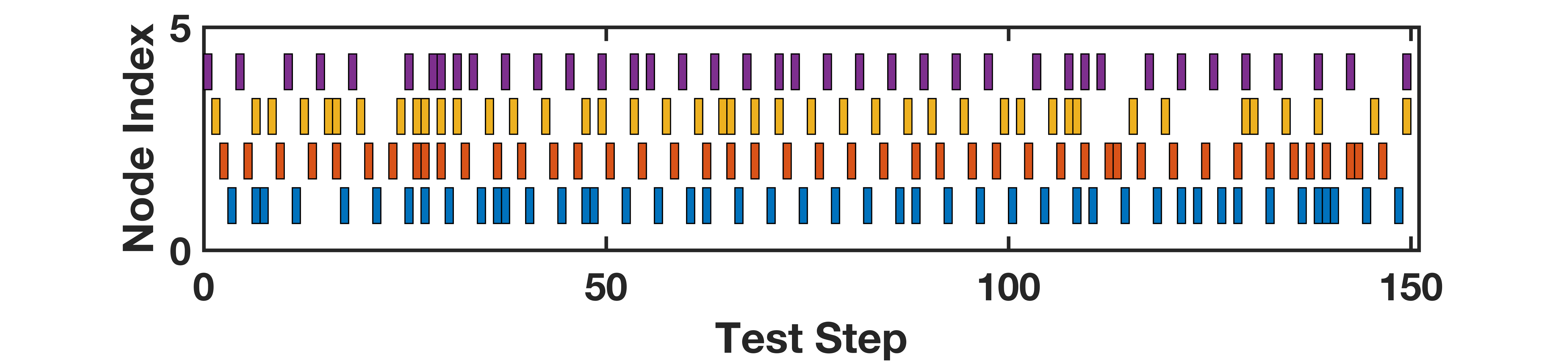}}	
    \vspace{-0.3cm}
    \subfigure[UANs with packet loss and observation completion
	\label{9c}]{\includegraphics[width=9cm]{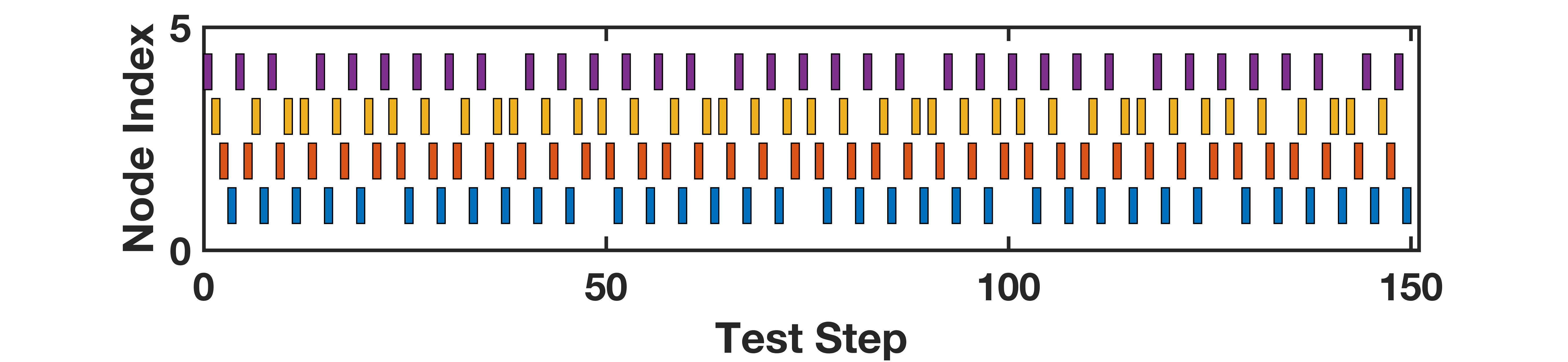}}
    
	\caption{Node transmission details under various test conditions in equidistant UANs. Each marker means that the node with the corresponding index sends a data packet in the current slot. } 
	\label{9}
\end{figure} 

\subsubsection{Trade off of throughput and fairness} We set the test step as 150, and this set of experiments does not have packet loss caused by packet error. As represented in Fig.~\ref{8}, we can observe that the trained EA-MAC adaptively balances throughput and fairness in both non-equidistant UANs and equidistant UANs. However, nodes in non-equidistant UANs transmit more packets than those in equidistant UANs. This is because nodes in equidistant UANs are unable to transmit concurrently due to high packet collisions, but nodes in non-equidistant UANs can employ the characteristic of long propagation delay to enable more concurrent transmissions. This demonstrates that the proposed DRL-based algorithm effectively captures and adapts to the distinct characteristics of different UAN topologies. In the 6-node non-equidistant UAN, the fifth sender can transmit packets concurrently with the other four senders. Under such a condition, the transmissions of node~5 do not compromise the fairness of data transmissions among the other senders; instead, they contribute to enhancing the overall network throughput. Therefore, node 5 transmits more packets than the other four senders, demonstrating the strong adaptability of EA-MAC to the network environment.

\subsubsection{Observation Completion} In this set of experiments, we use a 5-node equidistant UAN—characterized by complex collision scenarios and unstable factors—as an example to evaluate EA-MAC over 150 steps under various conditions. As represented in Table.~\ref{t3}, when the network occurs packet loss (we set 20\% data loss and 30\% ACK loss), although the fairness mechanism still plays a role, the lack of observation information can lead to inappropriate action decisions, resulting in a reduced number of packets received by the sink node. Further incorporate the details of Fig.~\ref{9b}, this is because some senders in such a condition decide to transmit data packets in the same slot, increasing data collisions. EA-MAC with an observation completion strategy can overcome the above issue. As shown in Table.~\ref{t3} and Fig.~\ref{9c}, EA-MAC is able to determine collision-free actions for each node even under packet loss conditions. This demonstrates the effectiveness of the proposed observation completion strategy in supporting reliable operation in real-world UANs.


\begin{figure}[t]
	\centering	
	\subfigure[Equidistant Topology\label{4a}]
    {\includegraphics[width=8cm]{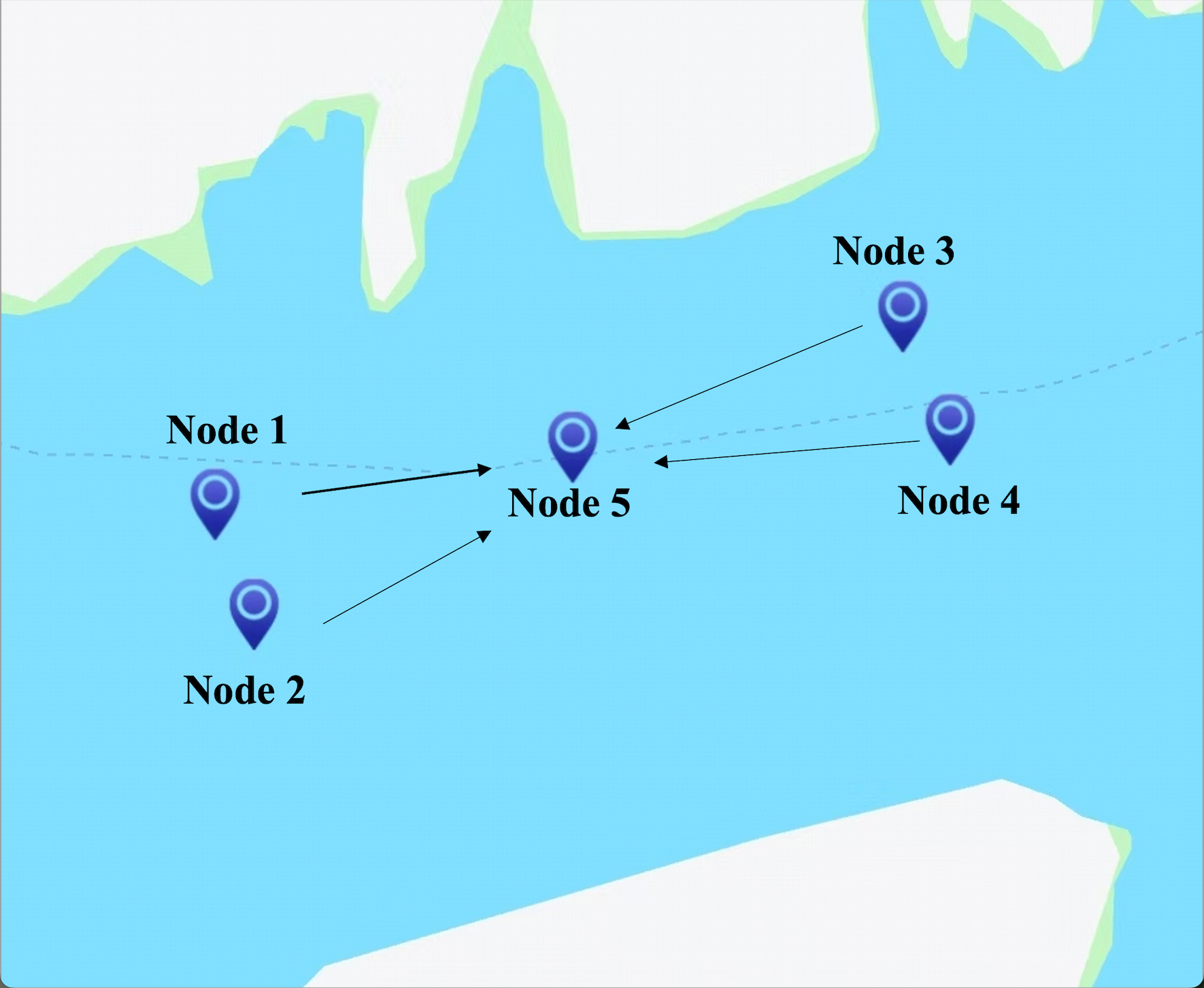}}

    \subfigure[Non-equidistant Topology
	\label{4c}]{\includegraphics[width=8.2cm]{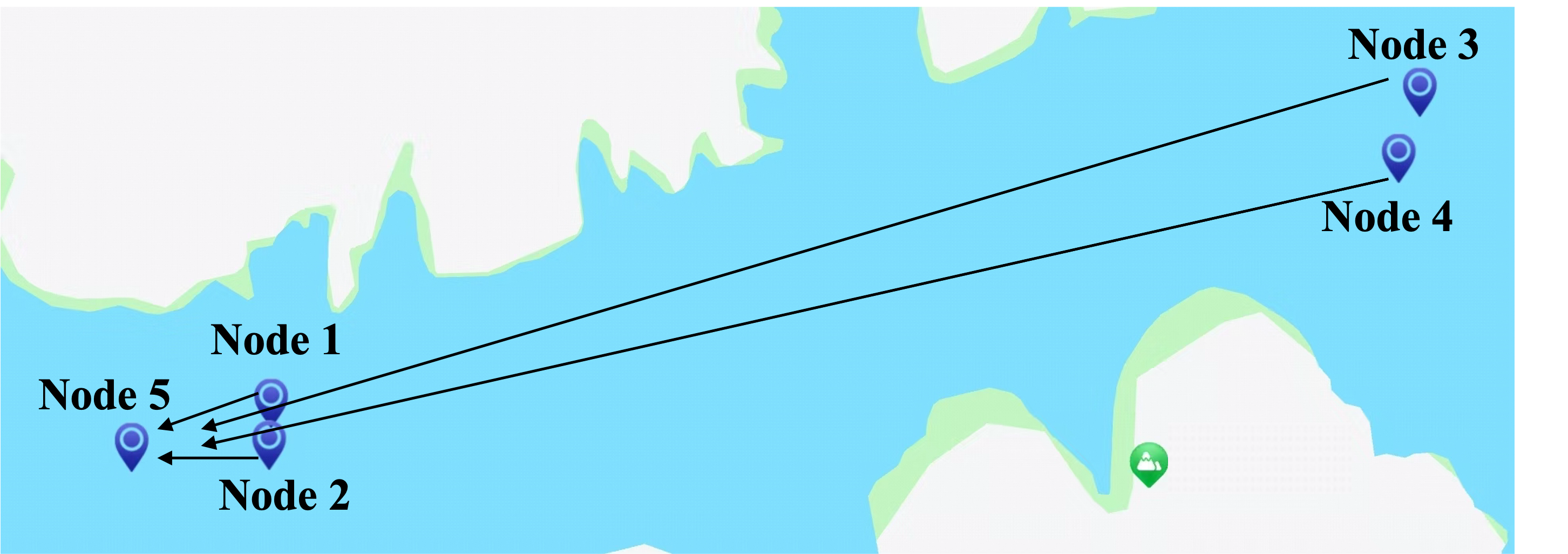}}
    
    \subfigure[Test Equipments
	\label{4d}]{\includegraphics[width=8cm]{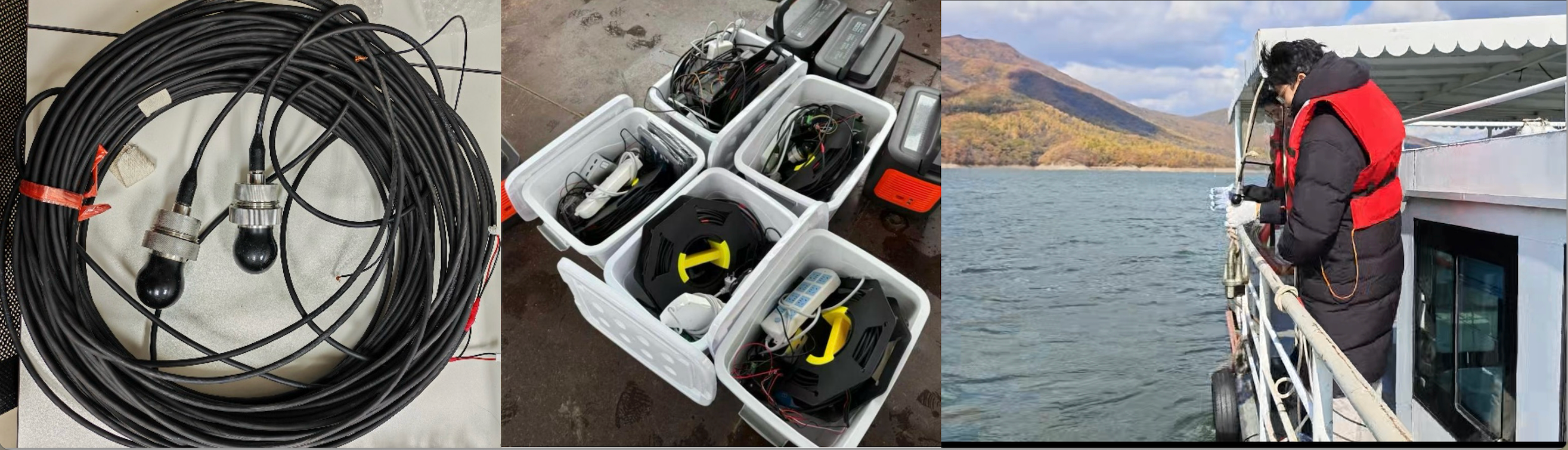}}
    
	\caption{A 5-node UAN deployed in Songhua Lake, Jilin, China, 2025. Node~1 to node~4 are senders, and node~5 is the sink node.} 
	\label{4}
\end{figure} 

\subsection{Real-World Field Experiments}

We deployed the trained network on real underwater acoustic modems and conducted actual field experiments (a 5-node network) at Songhua Lake in May and October 2025, as represented in Fig.~\ref{4}. EA-MAC and the baseline methods are evaluated in both equidistant and non-equidistant networks.

\subsubsection{Baseline Methods}

In real-world field experiments, we compare EA-MAC with the following protocols to evaluate its performance:

\begin{itemize}
    \item \textbf{Delayed-reward Deep-reinforcement Learning Multiple Access (DR-DLMA)} \cite{r5}: In DR-DLMA, only one node adopts DQN to determine its action: no transmission or transmission, while other nodes transmit packets with TDMA or ALOHA. Meanwhile, DR-DLMA assumes the reward can be exactly obtained after twice the one-way propagation delay following the agent’s action.    
    \item \textbf{Deep-reinforcement Learning-based Medium Access Control (DL-MAC)} \cite{r6}: DL-MAC still has only one node that adopts DQN like DR-DLMA. DL-MAC designs a time model, where the nodes are allowed to transmit packets with different delays in different time slots. Therefore, the agent has more actions to choose from: no transmission, or transmission with a choice of delay.   
    \item \textbf{Underwater ALOHA Q-learning (UW-ALOHA-Q)}~\cite{r22}: UW-ALOHA-Q employs Q-learning scheme to select a slot for each sender. Meanwhile, it retains random back-off of ALOHA. After a collision occurs, the node randomly delays the start time of its next frame according to a uniform distribution. 
    \item \textbf{ Slotted-Floor Acquisition Multiple Access (S-FAMA)} \cite{r12}: A classical handshake underwater MAC protocols. Before sending data packets, nodes should send control packets, called Request To Send (RTS) and Clear To Send (CTS), to reserve the channel. Only when the reservation is successful can nodes send data packets. 
    \item \textbf{TDMA and ALOHA}: We also select basic TDMA and ALOHA as representatives of traditional simple protocols to compare with other DRL-based protocols. 
\end{itemize}

\begin{table}[t]
\caption{Transmission and reception results of various protocols in the equidistant UAN}
\hspace*{-0.2cm}
\begin{threeparttable}
\begin{tabular}{lllllllllc}
\toprule
& \multicolumn{4}{c}{TC\tnote{1}}  & \multicolumn{4}{c}{RAC\tnote{2}} & RDC\tnote{3} \\ \cmidrule(lr){2-5} \cmidrule(lr){6-9}\cmidrule(lr){10-10}
Node Index      & 1        & 2       & 3       & 4       & 1       & 2        & 3       & 4       & 5                   \\ \midrule
EA-MAC         & 23       & 25      & 23      & 24      & 4       & 19       & 8       & 10      & 84                  \\
DR-DLMA$_{\text{(1T1A)}}$\tnote{4} & 7        & 0       & 25      & 49      & 0       & 0        & 8       & 12      & 55                  \\
DR-DLMA$_{\text{(1T)}}$\tnote{5}   & 13       & 16      & 35      & 25      & 2       & 3        & 9       & 4       & 49                  \\
DL-MAC$_{\text{(1T1A)}}$  &61       & 86     & 25       & 54       & 2       & 16        &2        &10     &72                     \\
DL-MAC$_{\text{(1T)}}$    &93        &97       &100       &25      &3         &40        &14       &0      &105                     \\
UW-ALOHA-Q         & 33        & 33      &34       &33       & 1       & 12        & 2       & 8       & 72                  \\
S-FAMA         & 3        & 7      & 3       & 2       & 0       & 5        & 1       & 1       & 13                  \\
TDMA           & 25       & 25      & 25      & 25      & 6       & 21       & 6       & 13      & 88                  \\
ALOHA          & 18       & 24      & 29      & 28      & 4       & 7        & 5       & 6       & 56                  \\ \bottomrule
\end{tabular}
\begin{tablenotes}
			\footnotesize
			\item[1] TC means Transmission Count of each sender.
			\item[2] RAC means the Count of self-ACKs Received by each sender.
            \item[3] RDC means the Count of Data Received by the sink.
            \item[4] 1T1A means that one node employs TDMA, one node employs ALOHA, and the remaining two nodes employ DRL-based MAC.
            \item[5] 1T means that one node employs TDMA, and the remaining three nodes employ DRL-based MAC.
		\end{tablenotes}
\end{threeparttable}
\label{t4}
\end{table}

\begin{table}[t]
\caption{Transmission and reception results of various protocols in the non-equidistant UAN}
\hspace*{-0.2cm}
\begin{threeparttable}
\begin{tabular}{lllllllllc}
\toprule
& \multicolumn{4}{c}{TC}  & \multicolumn{4}{c}{RAC} & RDC \\ \cmidrule(lr){2-5} \cmidrule(lr){6-9}\cmidrule(lr){10-10}
Node Index      & 1        & 2       & 3       & 4       & 1       & 2        & 3       & 4       & 5                   \\ \midrule
EA-MAC         & 48       & 49      & 50      & 49      & 47       & 48       & 39       & 48      & 186                  \\
DR-DLMA$_{\text{(1T1A)}}$ & 8        & 67       & 25      & 49      & 2       & 13        & 3       & 6      & 39                  \\
DR-DLMA$_{\text{(1T)}}$   & 59       & 13      & 7      & 25      & 27       & 0        & 1       & 3       & 45                  \\
DL-MAC$_{\text{(1T1A)}}$  &79       & 57     & 25       & 45       & 17       & 16        &4        &4     &41                     \\
DL-MAC$_{\text{(1T)}}$    &93        &63       &95       &25      &46         &18        &56       &1      &143                     \\
UW-ALOHA-Q  & 34        & 34       & 34       & 33       & 30       & 24        & 14       & 23       & 113                   \\ 
S-FAMA         & 8        & 12      & 6       & 4       & 8       & 11        & 3       & 4       & 28                  \\
TDMA           & 25       & 25      & 25      & 25      & 23       & 25       & 18       & 25      & 93                  \\
ALOHA          & 32       & 42      & 28      & 35      & 16       & 26        & 13       & 20       & 75                  \\ \bottomrule
\end{tabular}
\end{threeparttable}
\label{t5}
\end{table}

\begin{figure*}[t]
	\centering	
	\subfigure[EA-MAC \label{10a}]{\includegraphics[width=4.4cm,height=1.7cm]{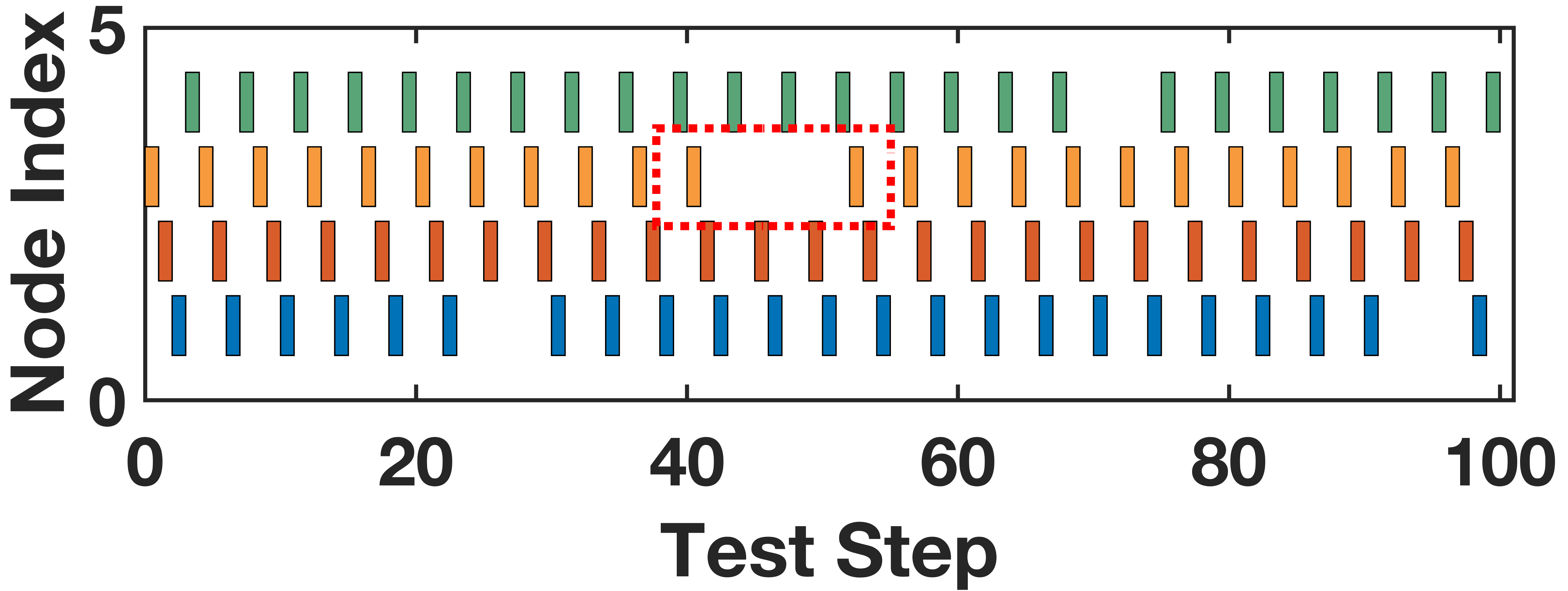}}   
	\subfigure[DR-DLMA (1T1A)
	\label{10b}]{\includegraphics[width=4.4cm,height=1.7cm]{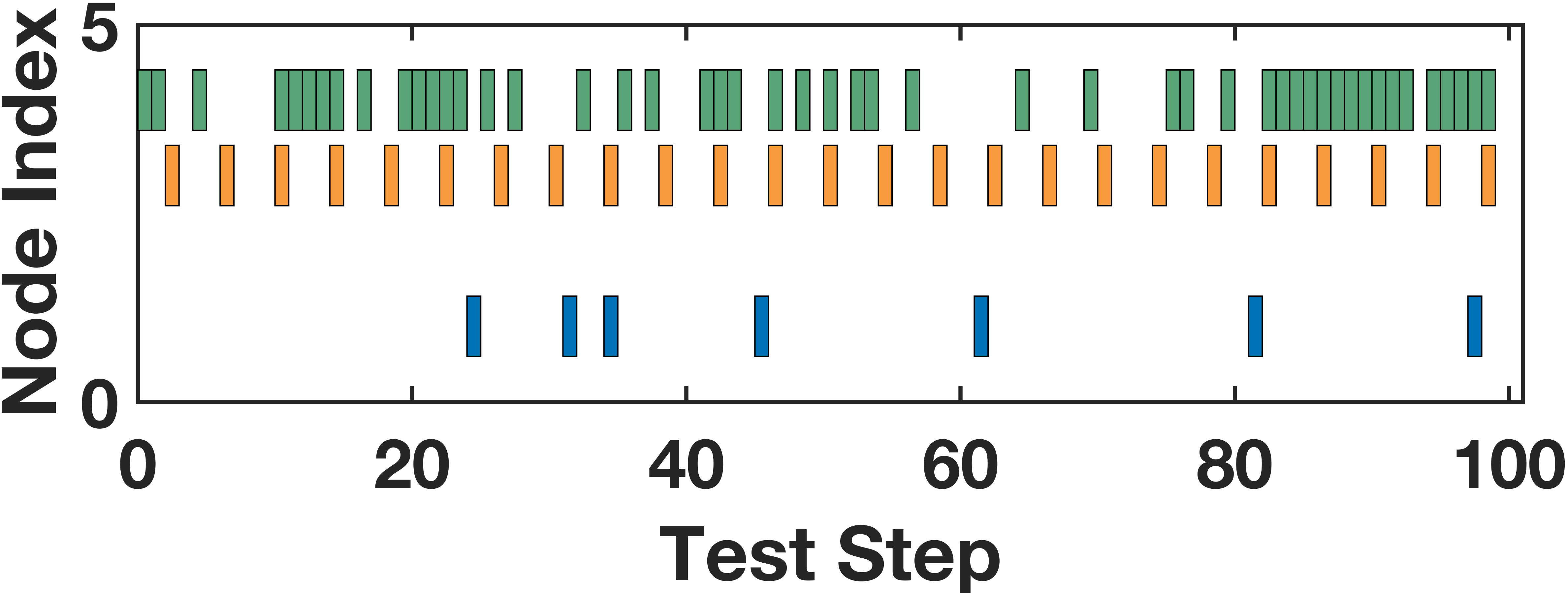}}	
    \subfigure[DL-MAC (1T1A)
	\label{10c}]{\includegraphics[width=4.4cm,height=1.7cm]{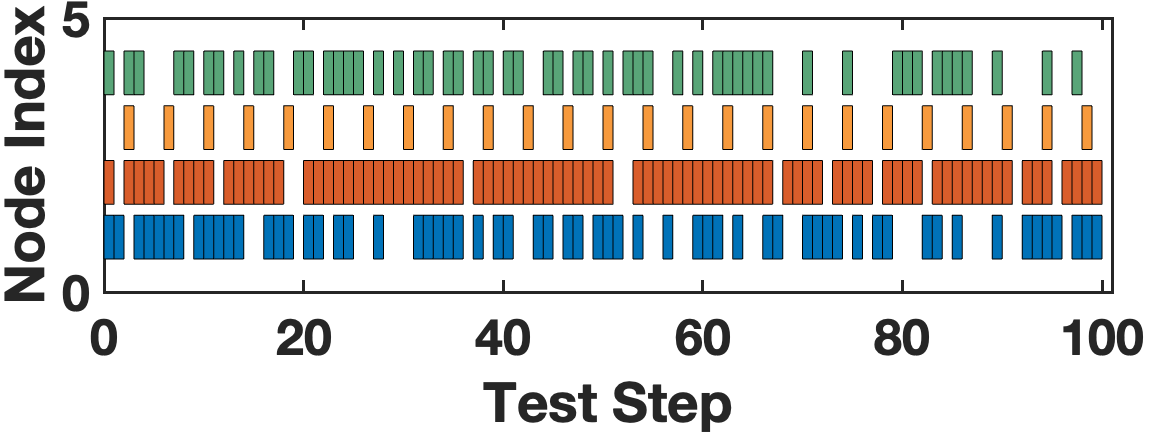}}
    \subfigure[UW-ALOHA-Q
	\label{10g}]{\includegraphics[width=4.4cm,height=1.7cm]{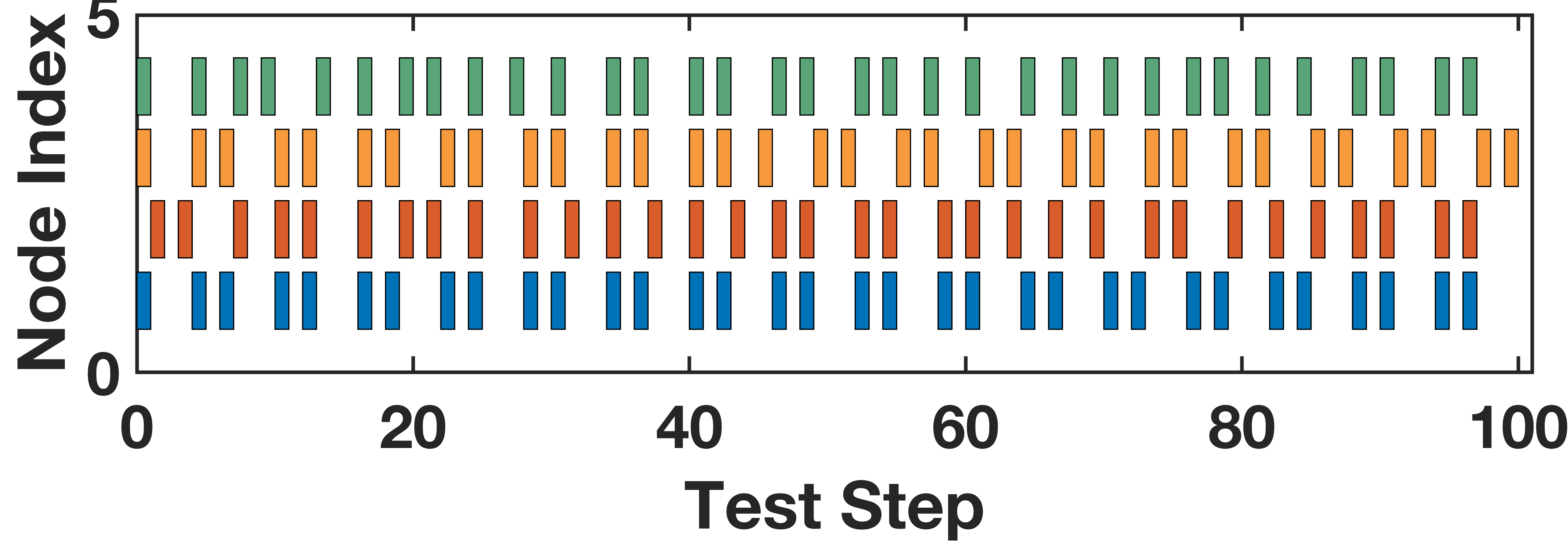}}
    
    \subfigure[S-FAMA \label{10d}]{\includegraphics[width=4.4cm,height=1.7cm]{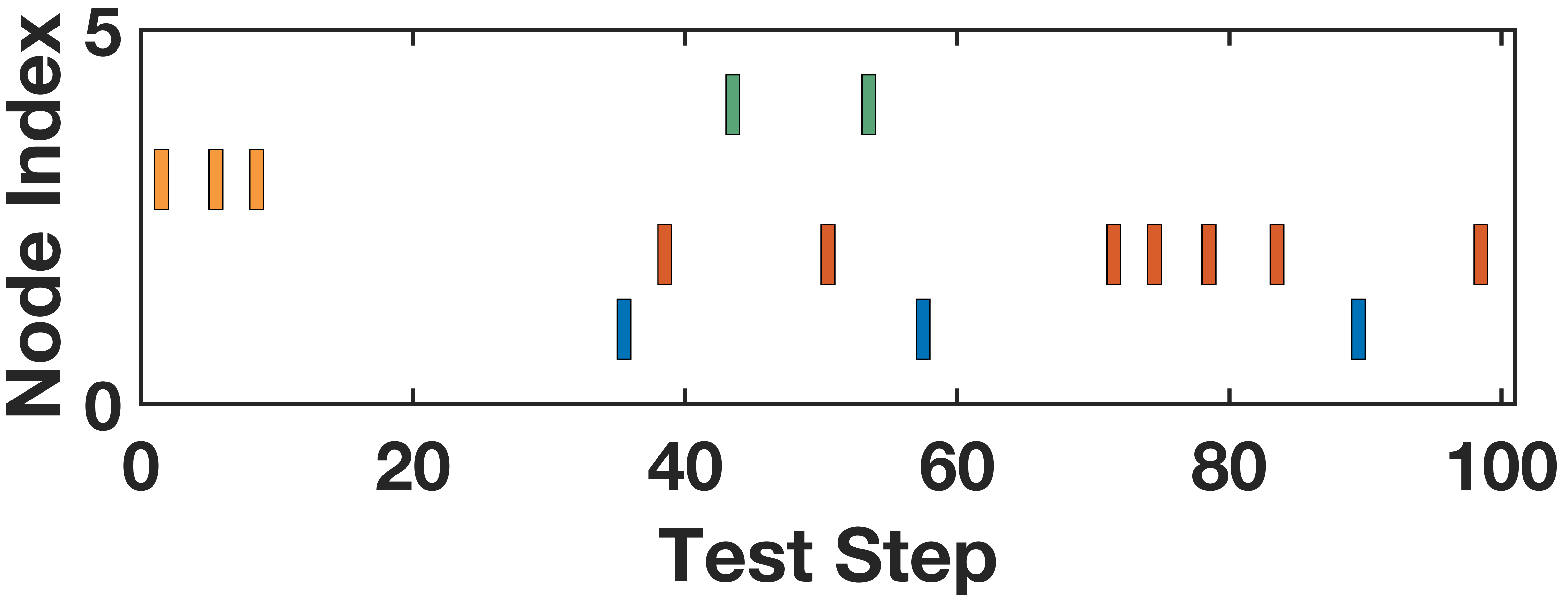}} 
    \hspace{0.02\textwidth}
	\subfigure[TDMA
	\label{10e}]{\includegraphics[width=4.4cm,height=1.7cm]{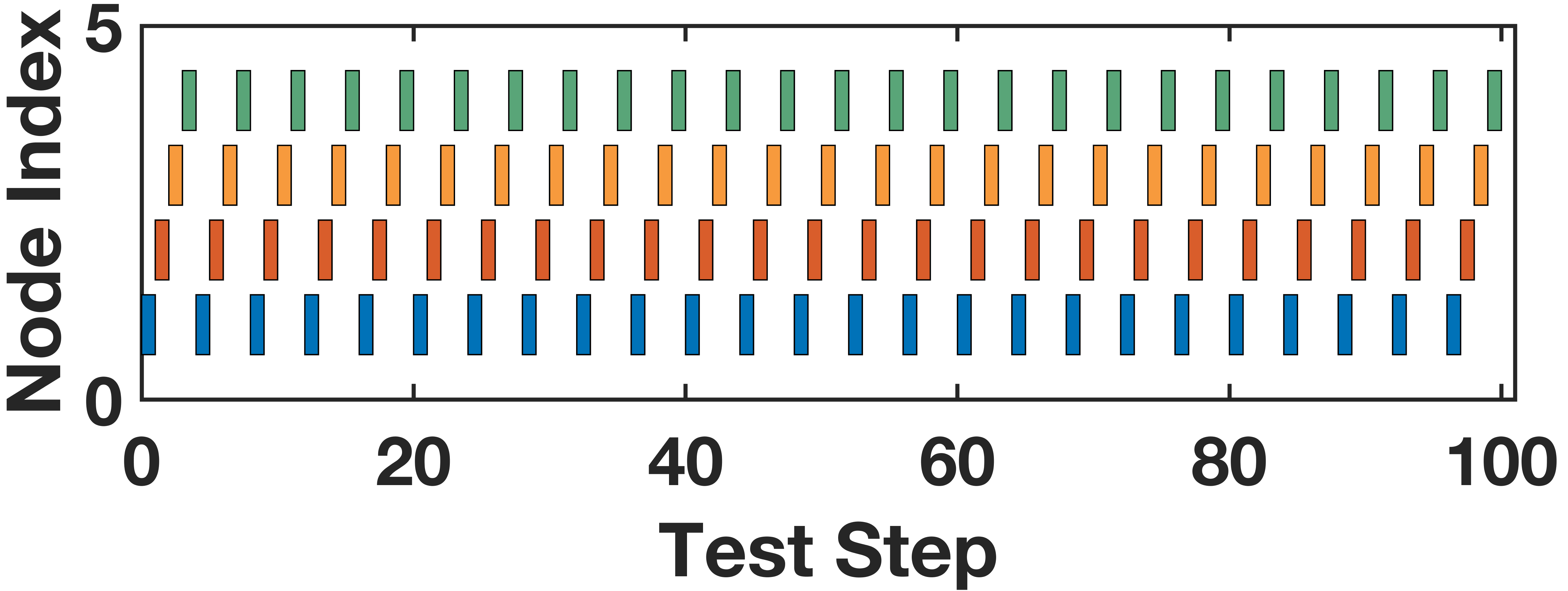}}
    \hspace{0.02\textwidth}
    \subfigure[ALOHA
	\label{10f}]{\includegraphics[width=4.4cm,height=1.7cm]{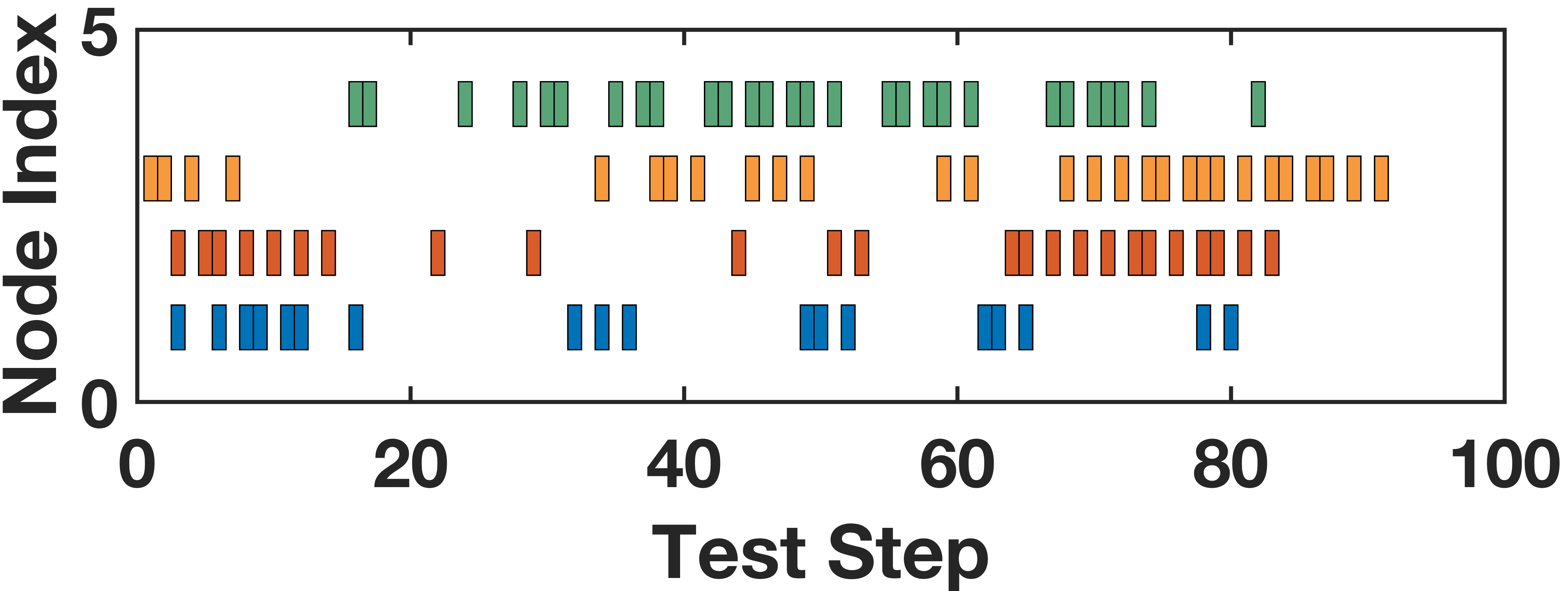}}
    
	\caption{Node transmission details of various MAC protocols in an equidistant UAN. Each marker means that the node with the corresponding index sends a data packet in the current slot. } 
	\label{10}
\end{figure*} 

\subsubsection{Performance Analysis of an Equidistant UAN}

In this set of experiments, we test all protocols for 100 steps. The transmission and reception results of each protocol are shown in Table.~\ref{t4}. We can observe that the RDC results are larger than the RAC results, which indicates the communication quality from the sink to senders is poorer than that from senders to the sink. This is in line with our previous experimental explorations as represented in Table.~\ref{t1}. 

In such a complex marine environment, the unreasonable assumptions of DR-DLMA and DL-MAC in reward acquisition delay lead to failure of intelligent decisions. We take DL-MAC (1T) as an example, nodes 1 to 3 are all agent nodes. Due to the fact that its trained network is unable to adapt to the uncertain delay and packet losses in the real environment, the transmission of the three nodes is completely irregular like ALOHA, as shown in Fig.~\ref{10c}. Although the sink can receive 105 packets, the reception ratio is only 0.33 lower than EA-MAC and TDMA. Although UW-ALOHA-Q achieves relatively fair packet transmissions among nodes,  this behavior mainly stems from its inherent protocol constraint, which requires each node to transmit once within every frame. Therefore, its fairness is enforced through a deterministic scheduling rule, rather than achieved through adaptive and intelligent decision-making. Meanwhile, UW-ALOHA-Q only relies ACK feedback to learn a collision-free transmission strategy. However, in real-world underwater environments, ACK loss can directly disrupt its Q-table update process, preventing nodes from maintaining a reasonable scheduling sequence. As a result, multiple nodes may select the same slot to transmit packets, leading to severe collisions.

For handshake-based protocols, the main issue is severe collisions among control packets. We take S-FAMA as an example in our real-world experiments. Compared with other protocols, nodes in S-FAMA send fewer data packets. This is because S-FAMA encounters high RTS/CTS collisions and losses in real UANs, resulting in a low successful ratio of channel reservation, and senders have very few opportunities to send data.

\begin{figure*}[t]
	\centering	
	\subfigure[EA-MAC \label{11a}]{\includegraphics[width=4.4cm,height=1.7cm]{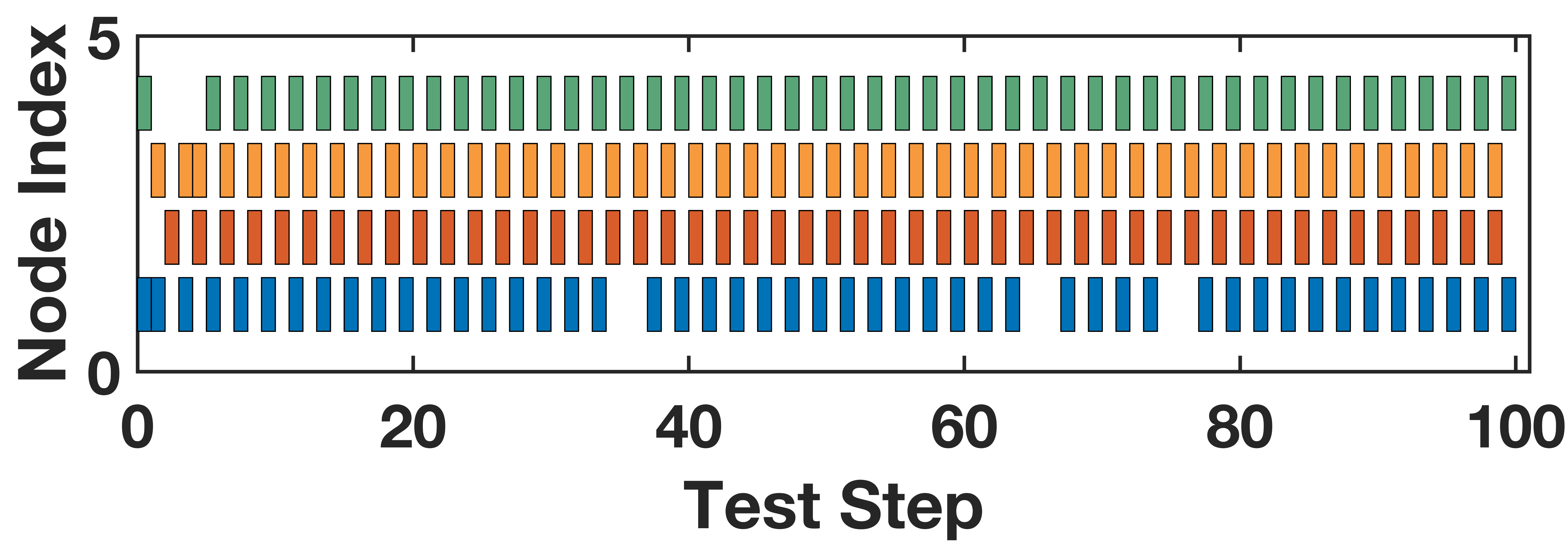}}   
	\subfigure[DR-DLMA (1T1A)
	\label{11b}]{\includegraphics[width=4.4cm,height=1.7cm]{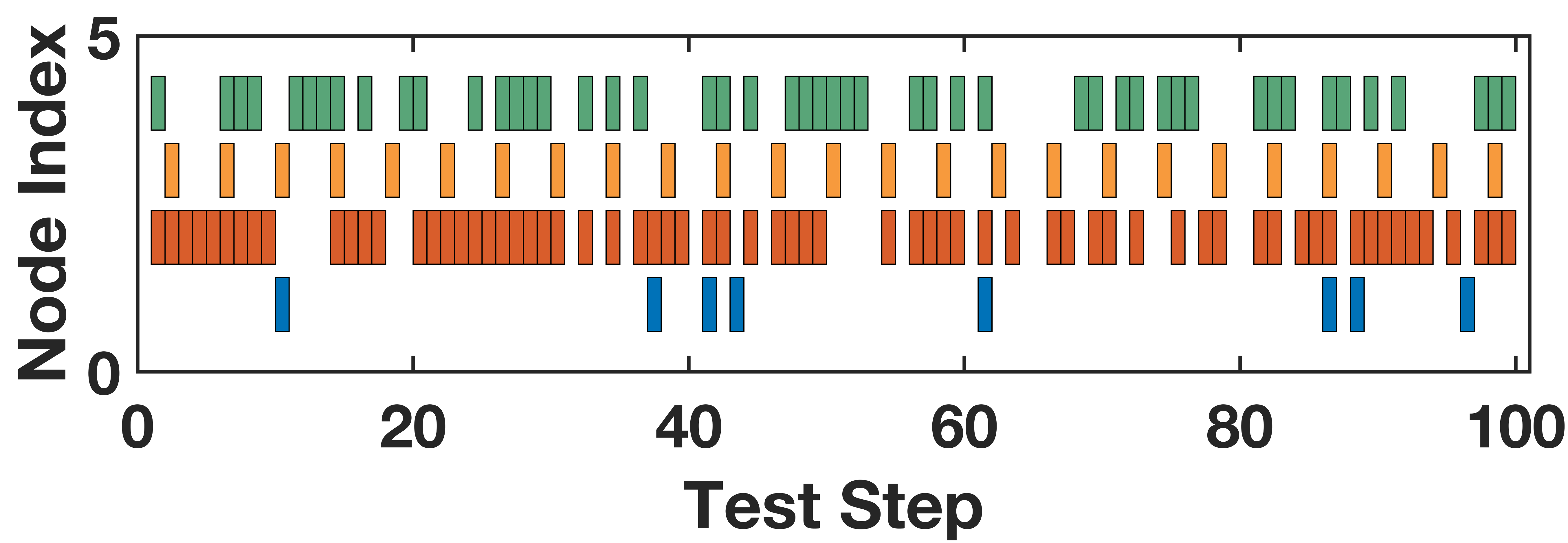}}	
    \subfigure[DL-MAC (1T1A)
	\label{11c}]{\includegraphics[width=4.4cm,height=1.7cm]{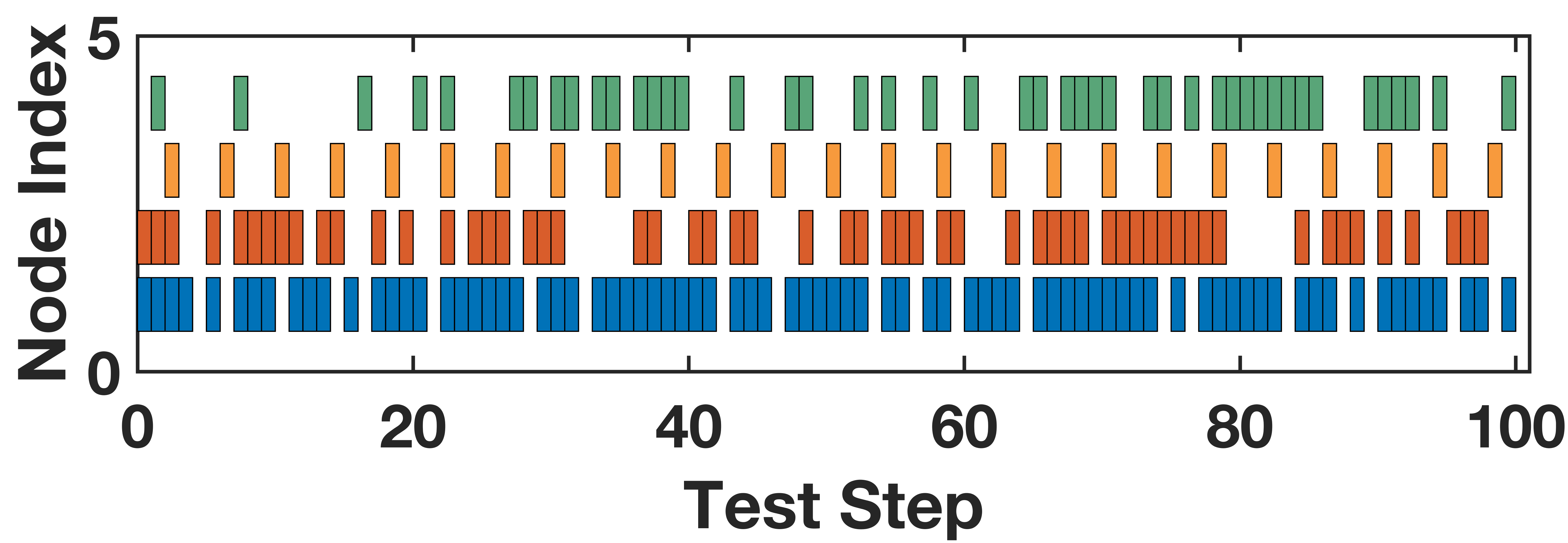}}
    \subfigure[UW-ALOHA-Q
	\label{11d}]{\includegraphics[width=4.4cm,height=1.7cm]{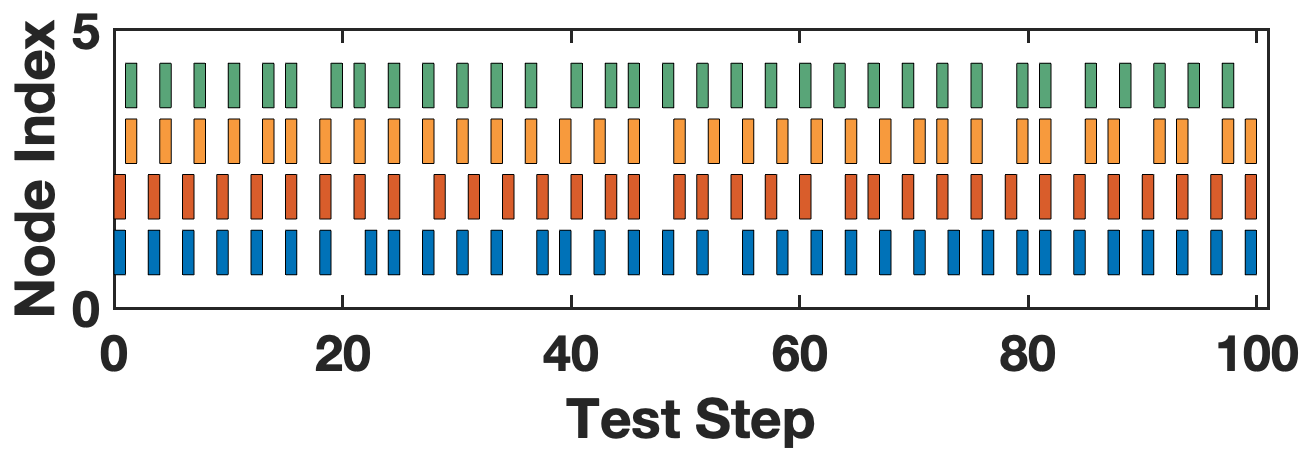}}
    
    \subfigure[S-FAMA \label{11e}]{\includegraphics[width=4.4cm,height=1.7cm]{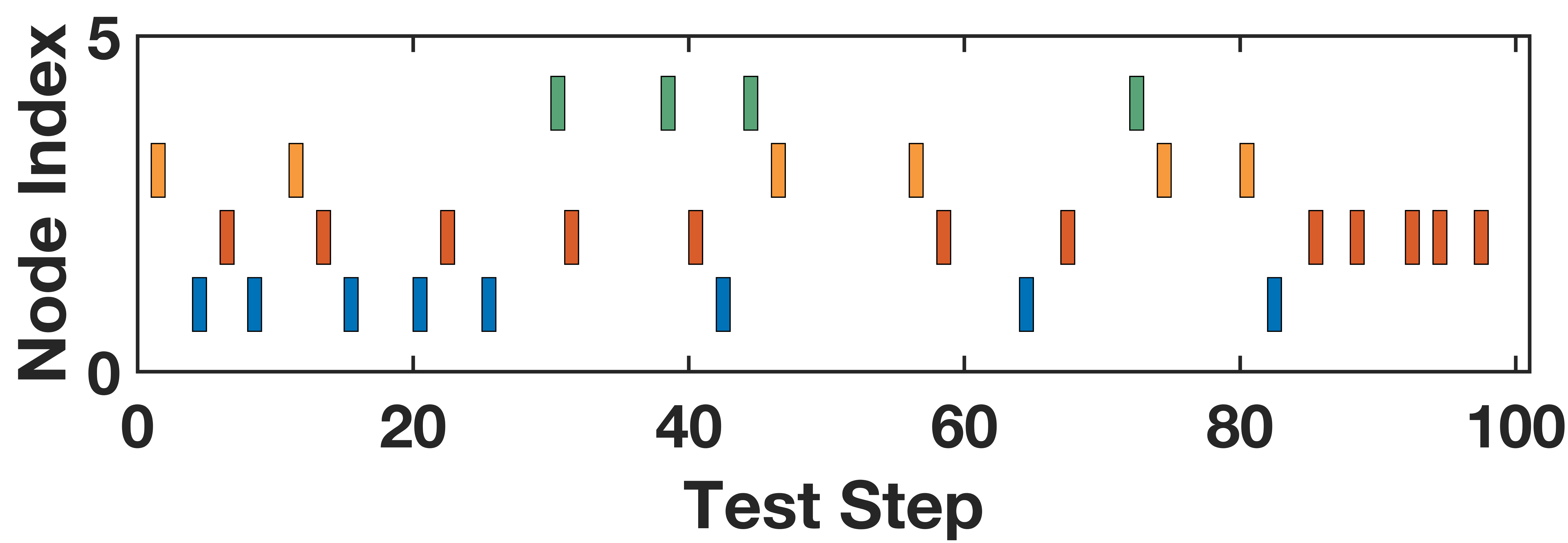}} 
    \hspace{0.02\textwidth}
	\subfigure[TDMA
	\label{11f}]{\includegraphics[width=4.4cm,height=1.7cm]{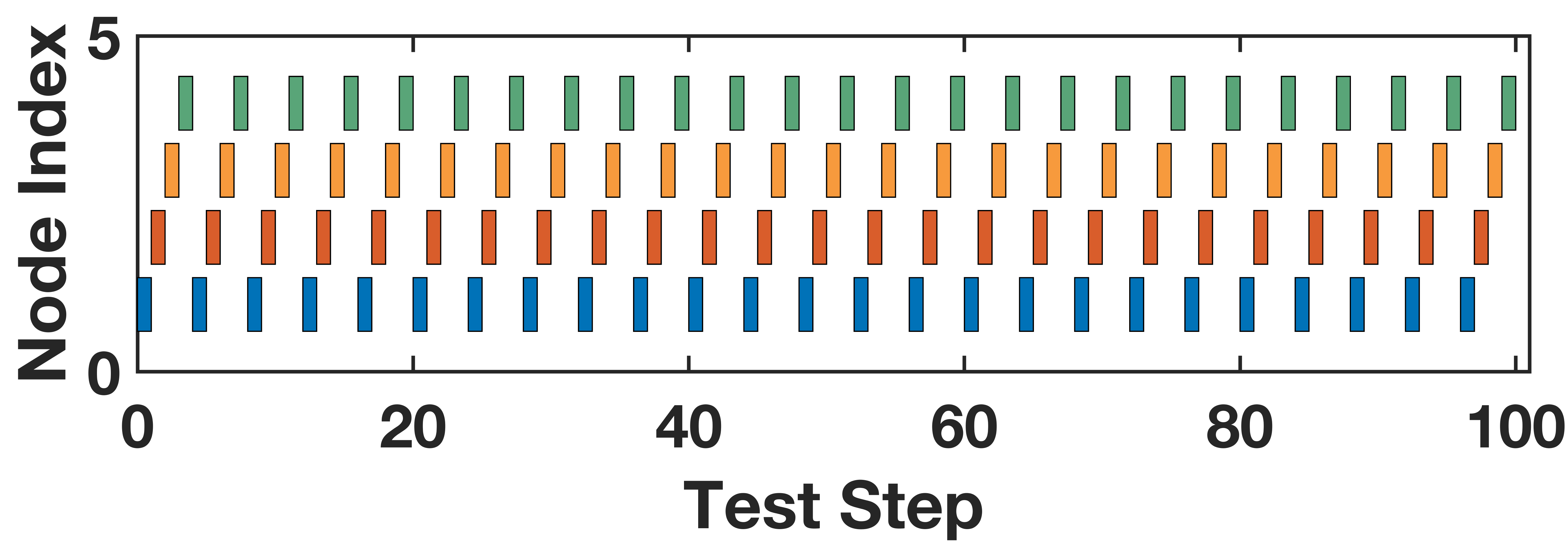}}
    \hspace{0.02\textwidth}
    \subfigure[ALOHA
	\label{11g}]{\includegraphics[width=4.4cm,height=1.7cm]{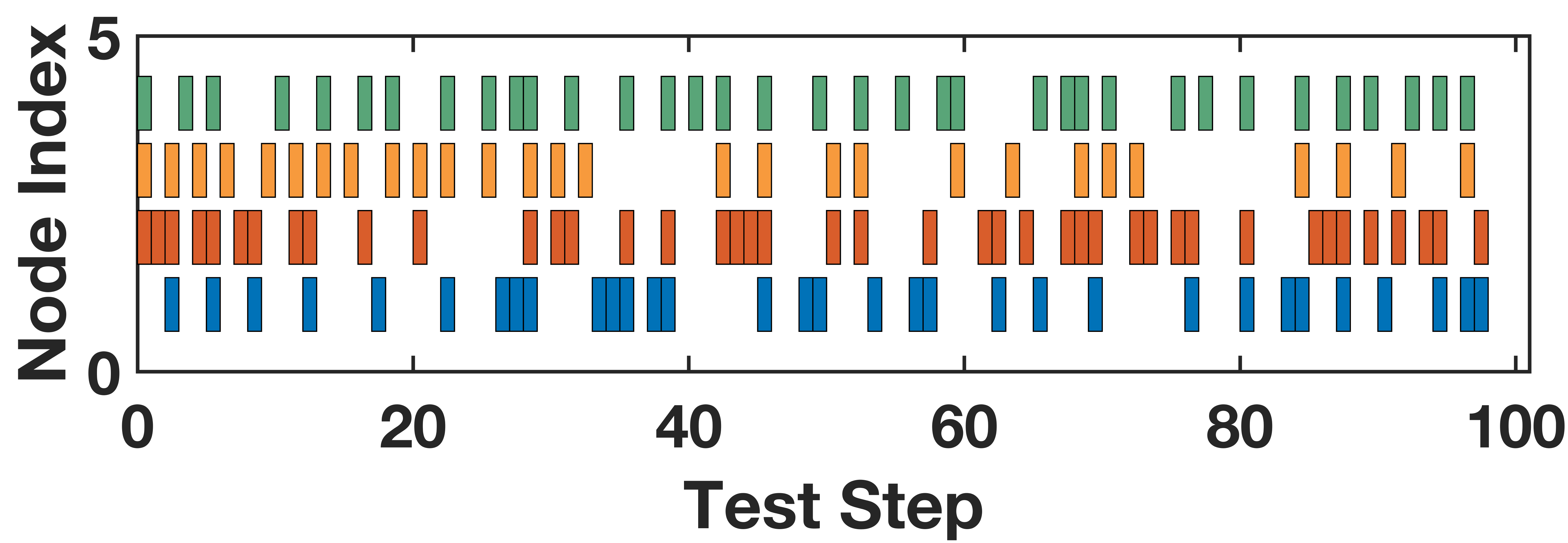}}
    
	\caption{Node transmission details of various MAC protocols in a non-equidistant UAN. Each marker means that the node with the corresponding index sends a data packet in the current slot. } 
	\label{11}
\end{figure*}

ALOHA is simple to implement, but has serious collisions. Our EA-MAC draws on the concept of its random access, expecting that nodes can adaptively decide their own scheduling action and avoid collisions without extra information exchange like other complex protocols. In our real-world UAN (an equidistant UAN), only one node is allowed to transmit in each time slot to avoid collisions. Based on the results of the Table.~\ref{t4}, Fig.~\ref{10a}, and Fig.~\ref{10e}, EA-MAC autonomously learns a TDMA-like transmission strategy with a random-access manner, which strongly demonstrates its effectiveness while simultaneously balancing protocol flexibility and overall network performance. However, EA-MAC is not equal to the TDMA protocol, it can improve concurrent transmissions in non-equidistant UANs, as shown in Fig.~\ref{8a} and Fig.~\ref{11a}. In addition, our real-world experiments occur continuous packet losses, and EA-MAC can enter a silence state to save energy. Take node~3 in Fig.~\ref{10a} as an example, it fails to receive any ACK packets from step 36 to 52. Based on our design, the node determines that the communication link is disconnected if it fails to receive any ACK packets for eight consecutive steps. Therefore, node~3 stops sending data packets from step 44 until it successfully receives ACK packets again at step 53. Experimental results demonstrate that EA-MAC can save more energy when the communication links are unstable. Meanwhile, such a flexible and autonomous design does not affect the subsequent decisions of the neural network.

\subsubsection{Performance Analysis of a Non-equidistant UAN}

Compared with the equidistant UAN, the non-equidistant UAN performs better performance within the same number of test steps, as represented in Table.~\ref{t5} and Fig.~\ref{11}. This is because the non-equidistant topology introduces differences in propagation distance among nodes, which provides opportunities for concurrent transmissions by different nodes due to long propagation delay of underwater environments, and thereby improves channel utilization.

However, the other baselines still perform worse than EA-MAC. Their unsatisfactory performance in the non-equidistant UAN can be attributed to the same reasons observed in the previous equidistant-UAN experiment: 1) conventional protocols (S-FAMA, TDMA, ALOHA) lack the adaptability to respond to environmental variations; 2) existing DRL-based protocols (DR-DLMA, DL-MAC, UW-ALOHA-Q) make overly strong assumptions about the underwater environment, failing to effectively cope with delayed reward feedback and missing observations in practical underwater scenarios. Since these limitations are fundamental rather than topology-specific, their poor performance persists in the non-equidistant UAN.

For EA-MAC, as most functionalities have already been validated in the previous experiment, we do not repeat the same analysis here. Instead, we focus on the concurrency performance to validate whether EA-MAC can adapt to underwater environmental variations and provide flexible access strategies accordingly. As represented in Fig.~\ref{11a}, EA-MAC allows a nearby node (node 1 or node 2) and a distant node (node 3 or node 4) to transmit packets in the same slot. This demonstrates that EA-MAC can capture environmental variations, learn the underlying collision relationships among nodes, and adapt its access decisions accordingly to best match the current scenario. Meanwhile, as shown in Table.~\ref{t5}, EA-MAC maintains a satisfied transmission fairness while supporting concurrent transmissions. This indicates that EA-MAC's adaptive suppression strategy not only regulates transmission fairness among nodes, but also preserves consistency among nodes with concurrency potential. In other words, for two nodes with concurrency potential, EA-MAC suppresses or promotes their transmissions simultaneously, thereby achieving a balance between fairness and throughput.

\section{Conclusion}

To enhance the flexibility and intelligence of underwater MAC protocols, we study DRL algorithms' practicality in real-world UANs. Through real field experiments, we analyze the application challenges of DRL in UANs, including uncertain reward acquisition delay, incomplete observations, and the trade-off between throughput and fairness. Based on the above challenges, we propose EA-MAC that considers long propagation delay, observation loss, and balances multiple reward factors to achieve efficient entire autonomous access in UANs. Experimental results demonstrate that EA-MAC allows each sender in real-world UANs to act as an intelligent agent, making its own transmission decisions independently, thereby achieving high-throughput and fair communication in a straightforward manner.

\section{Acknowledgment}

This work was supported in part by the National Natural Science Foundation of China under Grant 62501250 and Grant 62471201; in part by the Postdoctoral Science Foundation of China under Grant 2025M771509; in part by the Postdoctoral Fellowship Program of CPSF under Grant Number GZC20250178.

\bibliographystyle{IEEEtran}
\bibliography{ref}

\begin{IEEEbiography}
  [{\includegraphics[width=1in,height=1.25in,clip,keepaspectratio]{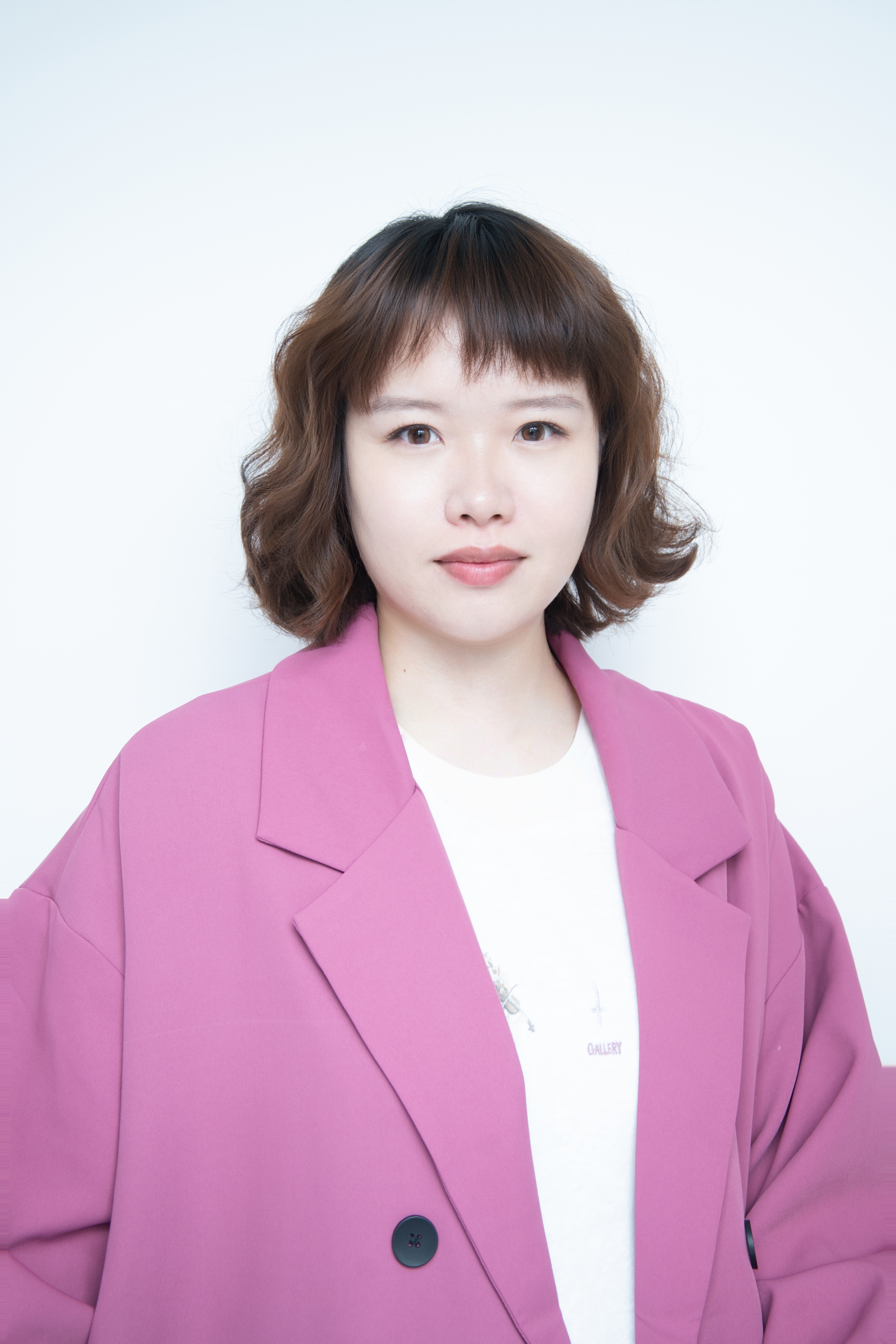}}]{Jiani Guo} received the BS degree (2016) in computer science and technology from Beijing Jiaotong University, Beijing, China, received PhD degree (2024) in Jilin University, Changchun, China. She is currently a Postdoctoral Researcher with the Department of Computer science and technology, Jilin University. Her current research interests include protocols design, performance analysis, and machine learning for underwater acoustic networks.
\end{IEEEbiography}

\begin{IEEEbiography}
  [{\includegraphics[width=1in,height=1.25in,clip,keepaspectratio]{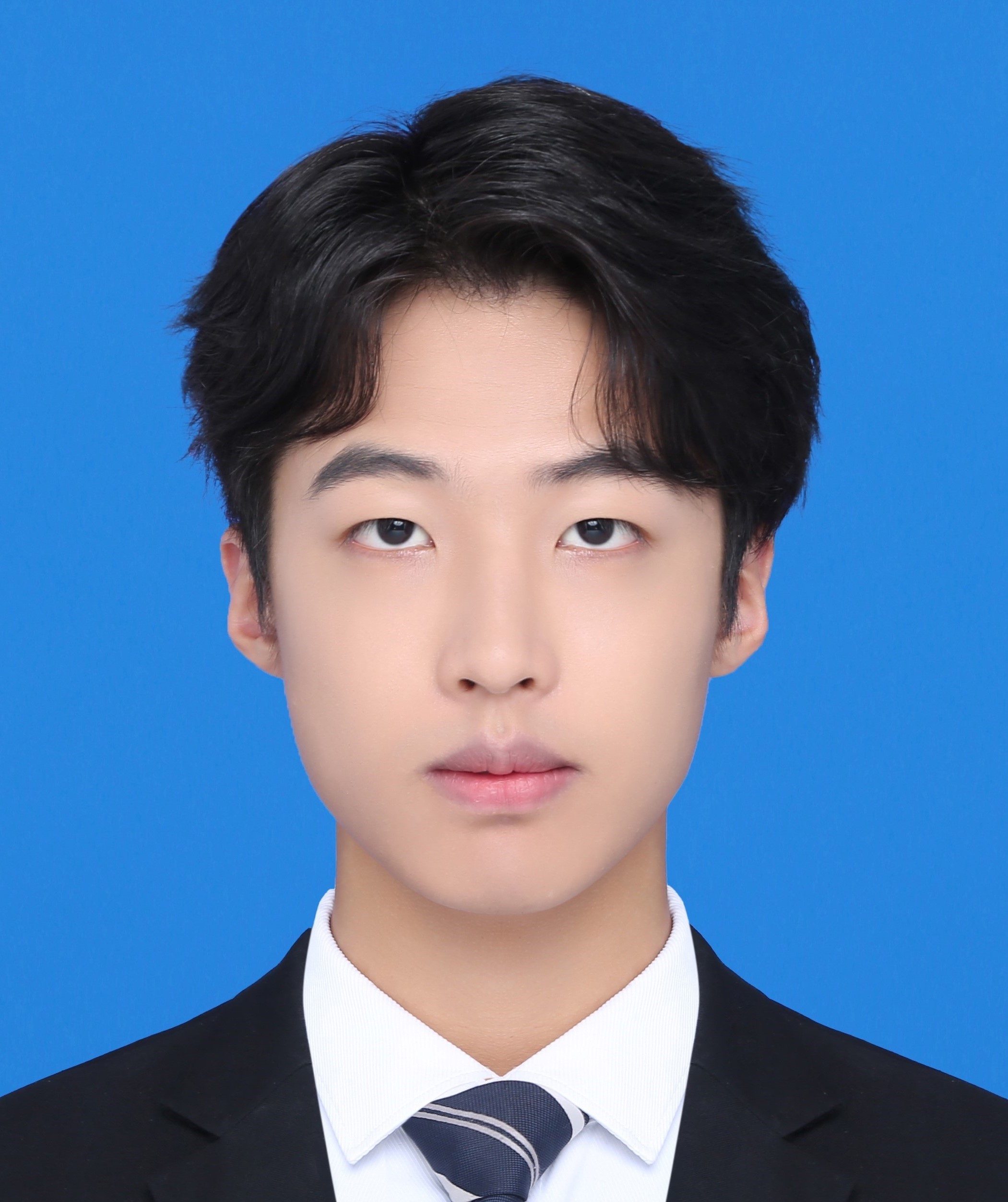}}]{Bingwen Huangfu} received the B.E. degree in computer science and technology from Jilin University, Changchun, China, in 2022. He is currently working toward the PhD degree at the College of Computer science and technology at Jilin University, Changchun, China.
  His current research interests include network architecture, resource allocation, and machine learning for underwater acoustic networks.
\end{IEEEbiography}

\begin{IEEEbiography}
  [{\includegraphics[width=1in,height=1.25in,clip,keepaspectratio]{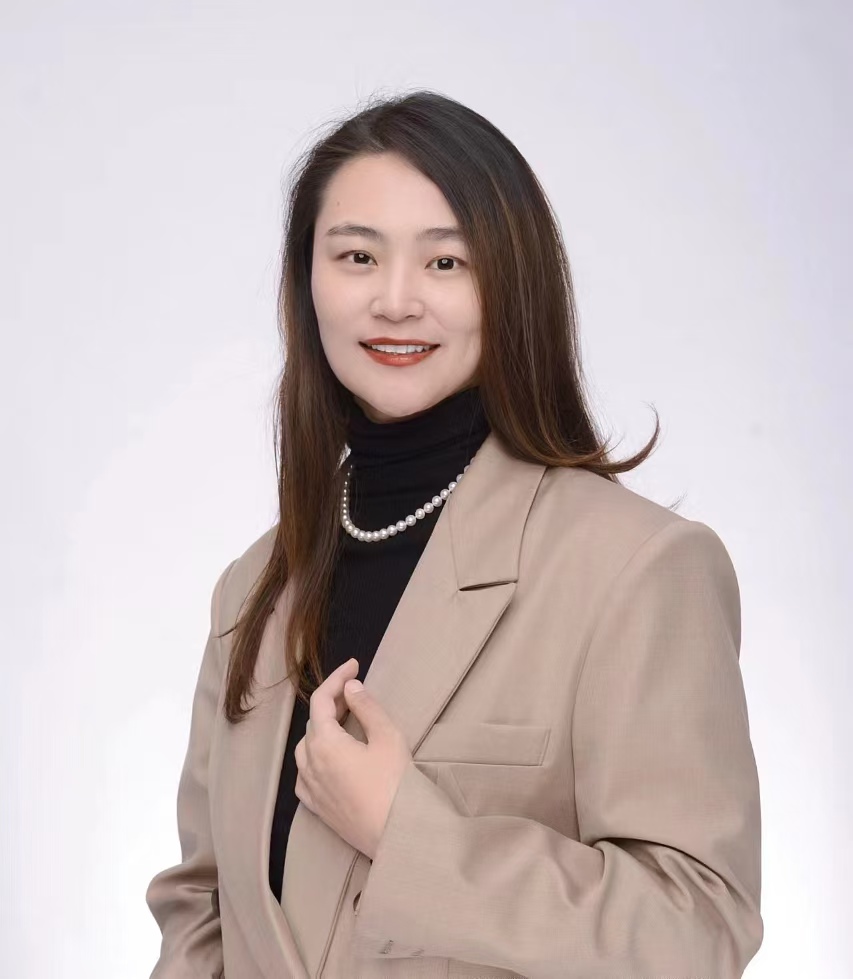}}]{Shanshan Song} (Member, IEEE) received the BS degree (2011) and MS degree (2014) in computer science and technology from Jilin University, China, received PhD degree (2018) in Management science and engineering from Jilin University, China. She was a Post-Doctoral Researcher with the Department of Computer science and technology, Jilin University, Changchun, China. She is currently an associate professor with the Department of Computer science and technology, Jilin University. Her major research focuses on underwater data collection, localization and navigation and machine learning. 
\end{IEEEbiography}

\begin{IEEEbiography}
  [{\includegraphics[width=1in,height=1.25in,clip,keepaspectratio]{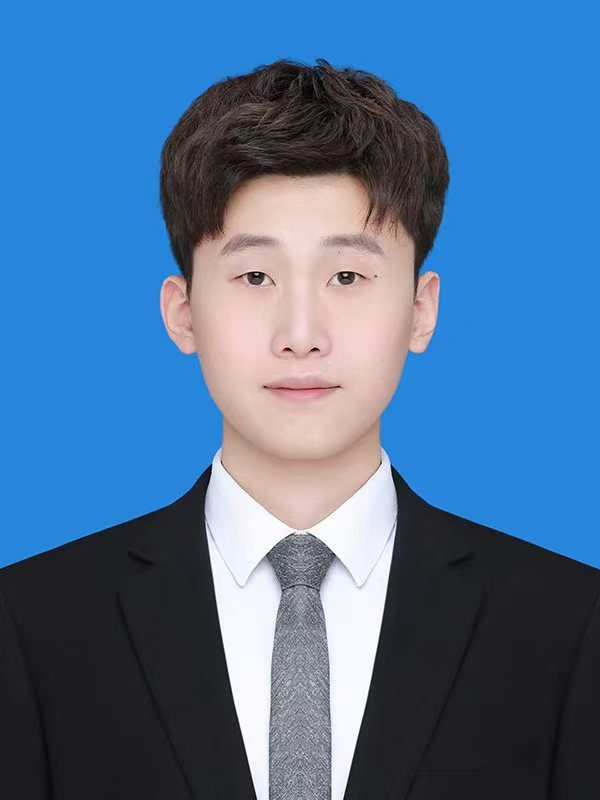}}]{Nan Sun} received the B.E. degree in software engineering from Yanshan University, Qinhuangdao, China, in 2023. He is currently working toward the M.S. degree with the College of Software Engineering, Jilin University, Changchun, China.
  His current research involves medium access control protocols for underwater acoustic networks.
\end{IEEEbiography}

\begin{IEEEbiography}
[{\includegraphics[width=1in,height=1.25in,clip,keepaspectratio]{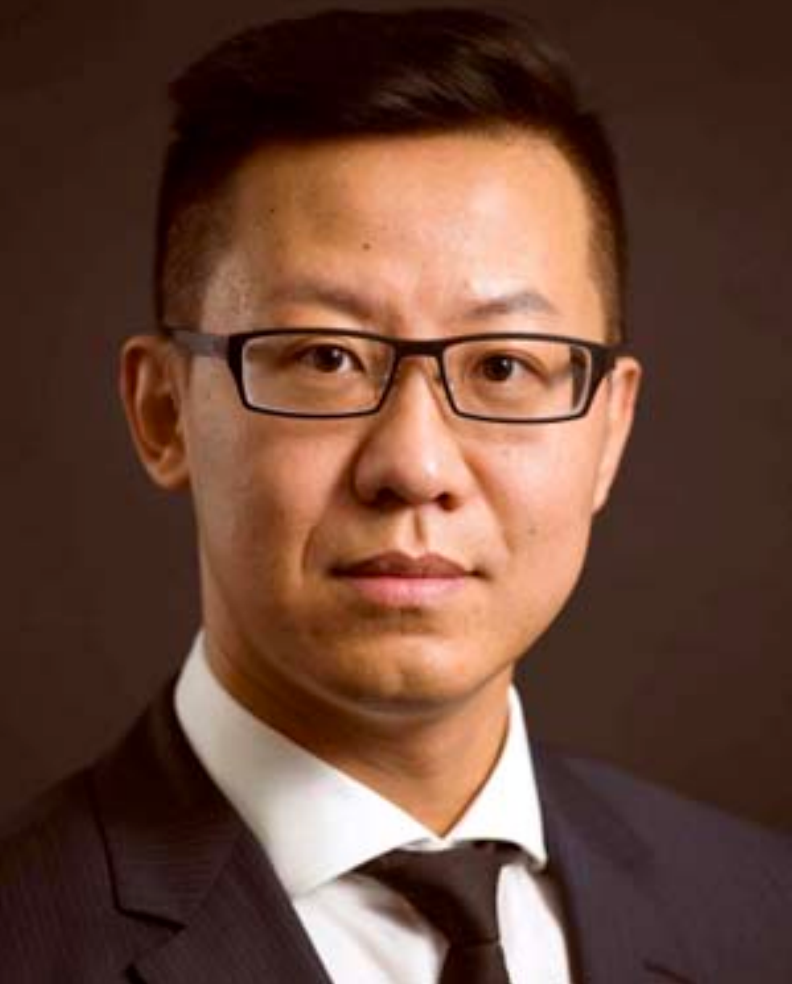}}]{Miao Pan} (Senior Member, IEEE) received the B.Sc. degree in electrical engineering from the Dalian University of Technology, Dalian, China, in 2004, the MASc. degree in electrical and computer engineering from the Beijing University of Posts and Telecommunications, Beijing, China, in 2007, and the Ph.D.degree in electrical and computer engineering from the University of Florida, Gainesville, FL, USA, in 2012. He is currently an Associate Professor with the Department of Electrical and Computer Engineering, University of Houston, Houston, TX, USA. His research interests include wireless/AI for AI/wireless, deep learning privacy, cybersecurity, and underwater communications and networking. He was the recipient of the NSF CAREER Award in 2014, IEEE TCGCC (Technical Committee on Green Communications and Computing) Best Conference Paper Awards 2019, and Best Paper Awards in ICC 2019, VTC 2018, Globecom 2017 and Globecom 2015, respectively. Dr. Pan is the Editor of IEEE OPEN JOURNAL OF VEHICULAR TECHNOLOGY, an Associate Editor for ACM Computing Surveys and IEEE INTERNET OF THINGS Journal (Area 5: Artificial Intelligence for IoT), and was an Associate Editor for IEEE INTERNET OF THINGS Journal (Area 4: Services, Applications, and Other Topics for IoT) from 2015 to 2018. He is also a Technical Organizing Committee for several conferences such as TPC Co-Chair for Mobiquitous 2019 and ACM WUWNet 2019.
\end{IEEEbiography}

\begin{IEEEbiography}
[{\includegraphics[width=1in,height=1.25in,clip,keepaspectratio]
{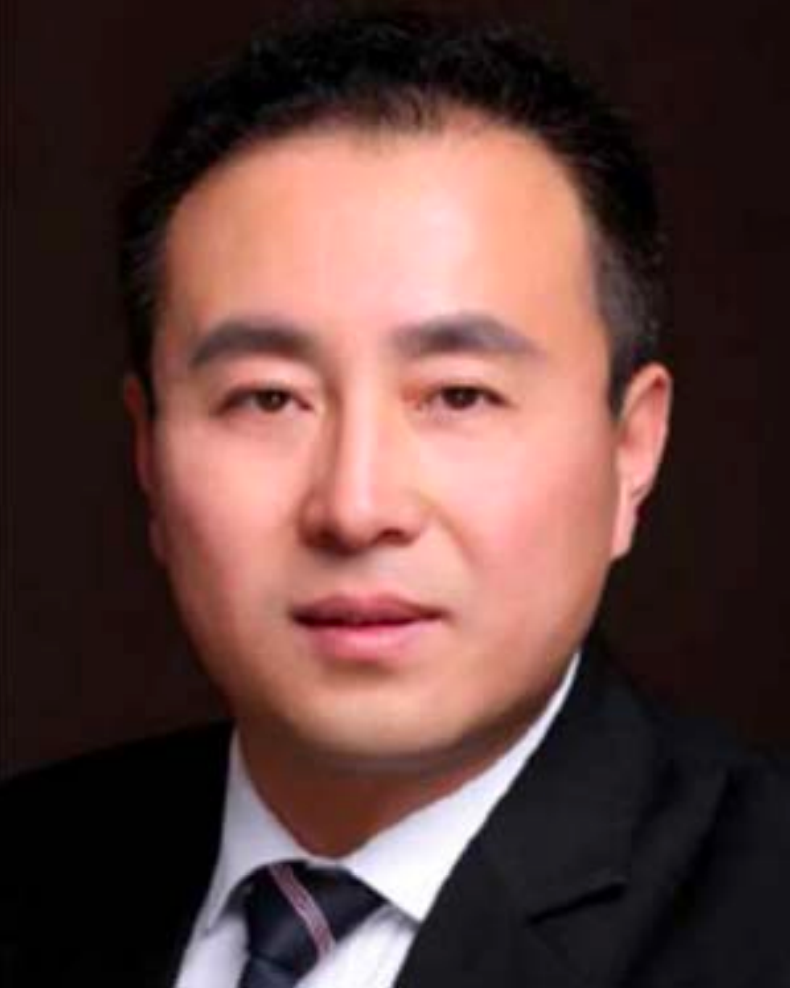}}]{Guangjie Han} (Fellow, IEEE) is currently a Professor with the Department of Internet of Things Engineering, Hohai University, Changzhou, China. He received his Ph.D. degree from Northeastern University, Shenyang, China, in 2004. In February 2008, he finished his work as a Postdoctoral Researcher with the Department of Computer Science, Chonnam National University, Gwangju, Korea. From October 2010 to October 2011, he was a Visiting Research Scholar with Osaka University, Suita, Japan. From January 2017 to February 2017, he was a Visiting Professor with City University of Hong Kong, China. From July 2017 to July 2020, he was a Distinguished Professor with Dalian University of Technology, China. His current research interests include Internet of Things, Industrial Internet, Machine Learning and Artificial Intelligence, Mobile Computing, Security and Privacy. Dr. Han has over 500 peer-reviewed journal and conference papers, in addition to 160 granted and pending patents. Currently, his H-index is 81 and i10-index is 381 in Google Citation (Google Scholar). The total citation count of his papers raises above 22300 times. 

Dr. Han is a Fellow of the UK Institution of Engineering and Technology (FIET). He has served on the Editorial Boards of up to 10 international journals, including the IEEE TII, IEEE TCCN, IEEE TVT, IEEE TNSM, IEEE Systems, etc. He has guest-edited several special issues in IEEE Journals and Magazines, including the IEEE JSAC, IEEE Communications, IEEE Wireless Communications, Computer Networks, etc. Dr. Han has also served as chair of organizing and technical committees in many international conferences. He has been awarded 2020 IEEE Systems Journal Annual Best Paper Award and the 2017-2019 IEEE ACCESS Outstanding Associate Editor Award. He is a Fellow of IEEE.
\end{IEEEbiography}

\end{document}